\pdfoutput=1

\documentclass[11pt]{article}
\usepackage[table]{xcolor}
\definecolor{verylightgray}{gray}{0.82}
\usepackage[]{acl}

\usepackage{times}
\usepackage{latexsym}
\usepackage{multirow}
\usepackage{adjustbox}
\usepackage{booktabs} 
\usepackage{subcaption}
\usepackage[T1]{fontenc}
\usepackage[utf8]{inputenc}
\usepackage{microtype}
\usepackage{inconsolata}
\usepackage{graphicx}
\usepackage{ccicons}
\usepackage[most]{tcolorbox}
\usepackage{tabularx}

\usepackage{xcolor}
\definecolor{posgreen}{RGB}{0,128,0}

\tcbset{
    promptbox/.style={
        colback=gray!20, 
        colframe=gray!50, 
        boxrule=0.5pt,
        arc=3pt,
        left=5pt,
        right=5pt,
        top=5pt,
        bottom=5pt,
        sharp corners,
        fontupper=\ttfamily\tiny, 
    }
}

\title{\textit{\MakeUppercase{J\scalebox{.75}{ump}}\MakeUppercase{S\scalebox{.75}{tarter}}}: Human-AI Planning with Task-Structured Context Curation}

\author{
 \textbf{Xuanming Zhang\textsuperscript{*}\textsuperscript{1}},
 \textbf{Sitong Wang\textsuperscript{*}\textsuperscript{1}},
 \textbf{Jenny Ma\textsuperscript{1}},
 \textbf{Alyssa Hwang\textsuperscript{2}}, \\
 \textbf{Zhou Yu\textsuperscript{1}},
 \textbf{Lydia B. Chilton\textsuperscript{1}}
\\
\\
 \textsuperscript{1}Columbia University,
 \textsuperscript{2}University of Pennsylvania
\\
\small{
 \texttt{
   \{xz2995, sw3504\}@columbia.edu, chilton@cs.columbia.edu
 }}
}

\begin{document}
\maketitle
\begingroup\def\thefootnote{*}\footnotetext{These authors contributed equally to this work.}\endgroup
\begin{abstract}
Human-AI collaboration on complex planning goals is bottlenecked by how LLM interfaces handle context: users must manually curate and re-surface relevant information across long and unstructured chat histories. 
Despite advances in long-context prompting and memory-augmented retrieval, this burden remains unresolved: users still have to identify and supply the right context at each decision point, regardless of how much the model can store or surface.
We propose \textit{JumpStarter}, a system that enables LLMs to collaborate with humans on complex goals by dynamically decomposing tasks to help users manage context. We specifically introduce \textit{task-structured context curation}, a framework that breaks down a user's goal into a hierarchy of actionable subtasks and scopes context to localized decision points, enabling finer-grained personalization and reuse. The framework is realized through three core mechanisms: \textit{context elicitation}, \textit{selection}, and \textit{reuse}. 
In a within-subjects user study, plans produced with \textit{JumpStarter} were rated substantially higher in quality than those produced with ChatGPT.
A complementary automatic simulation study shows that \textit{JumpStarter} consistently outperforms ChatGPT baselines, planning and memory agents, and workflow ablations. These findings show that effective human-AI planning depends not on the volume of context provided, but on attaching the right context to the right subtask at the right time.\footnote{Our code and data are available at \url{https://github.com/BillyZhang24kobe/jumpstarter}.}

\end{abstract}

\section{Introduction}

\begin{figure*}[h]
\centering
\includegraphics[width=1\textwidth]{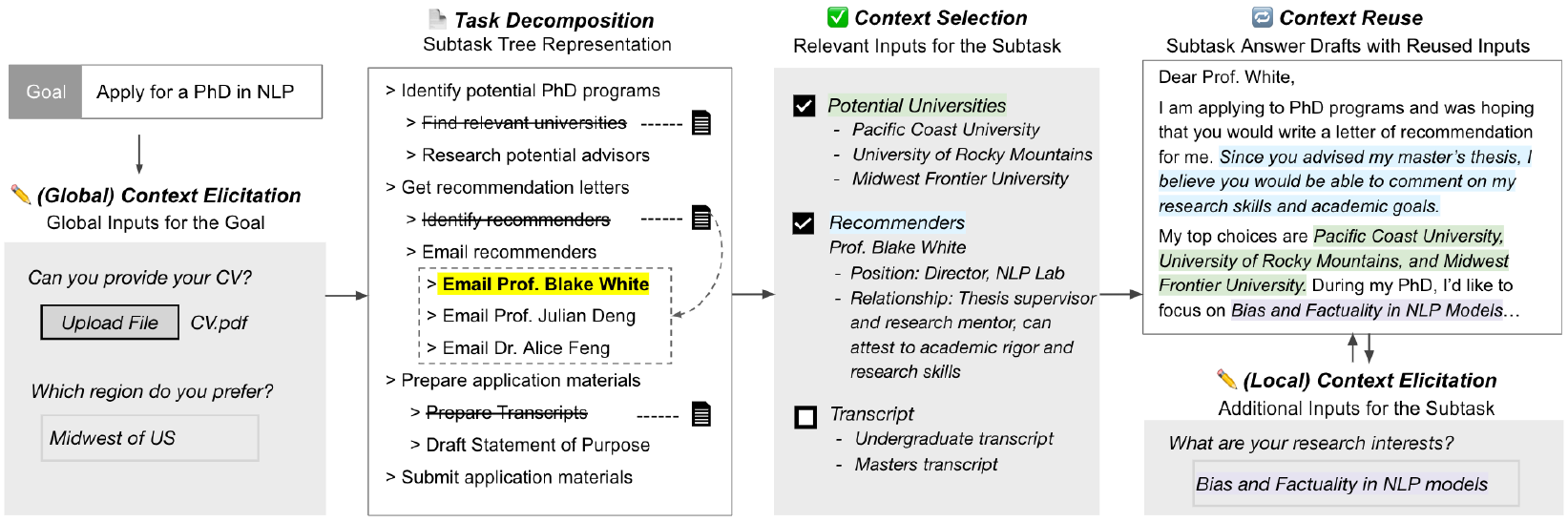}
\caption{\textit{JumpStarter} helps users get started on their personal goals through \textit{task-structured context curation}. It first takes the user's goal and elicits context for the goal. It then decomposes the goal into actionable subtasks. For each subtask, it helps users select relevant context and write answer drafts. It also aids users in refining these drafts by eliciting further context. \textit{Task-structured context curation} improves plan quality over ablations. Our user study showed that plans produced with \textit{JumpStarter} were rated substantially higher in quality than those produced with \texttt{GPT-4o} via the ChatGPT interface.}
\label{fig:teaser}
\end{figure*}

Planning is a core cognitive process for solving complex,  
goal-oriented tasks \cite{miller2021multi,ho2022people}. 
Recent advances in the planning abilities of large language models (LLMs) \cite{valmeekam2023planbench,shinn2023reflexion} have enabled human-AI planning across various domains, such as travel \cite{xie2024travelplanner}, manufacturing, and healthcare \cite{lee2025veriplan}. 
In these settings, LLMs assist users in decomposing complex goals into actionable steps \cite{wei2022chain, shinn2023reflexion}. 
However, while LLMs can support planning at scale, they still struggle with maintaining context over long user interactions \cite{jiang2023llmlingua}. 
Despite improvements in memory mechanisms and extended context windows \cite{luo2025large}, users must actively manage what information to provide, as LLMs frequently forget key details, leading to a degraded user experience. 
To generate personalized plans, users often resort to intensive prompt engineering \cite{zamfirescu2023johnny,sarkar2025conversational}, manually curating chat histories to elicit relevant responses, resupply information, and manage subtasks~\cite{liang2024c5}. 
This process is opaque, as users have little insight into which context is retrieved.

These problems share a common pattern: current LLM interfaces treat the conversation as a single, unstructured context, leaving users to decide what information is relevant at each stage of planning. LLM-based systems instead need to bind context to task structure: scoping it to specific subtasks, surfacing what is used at each step, and reusing intermediate progress as the plan evolves.

To address these limitations, we present \textit{JumpStarter}, a human-AI planning system built around a framework we call \textit{task-structured context curation}. Rather than treating the conversation as a single and unstructured context, \textit{JumpStarter} decomposes a user's goal into a hierarchy of actionable subtasks and binds context to each subtask, so that personalization and reuse happen at the level of individual planning decisions. \textit{JumpStarter} realizes this idea through three core mechanisms: \textit{context elicitation}, where the system asks targeted questions to fill in information missing for a specific subtask; \textit{context selection}, where the system and the user together surface only the prior inputs relevant to the current subtask; and \textit{context reuse}, where user-approved intermediate artifacts (e.g., emails, study schedules, etc.) are saved and carried forward into later subtasks. As illustrated in Figure~\ref{fig:teaser}, a user enters a goal, JumpStarter elicits relevant context, decomposes the goal into subtasks, and helps the user produce personalized drafts that become reusable context for later steps. Together, these mechanisms let users focus on one actionable step at a time while the LLM operates on localized, context-rich prompts, an alternative to labor-intensive long-context prompting that yields more structured, adaptive, and user-controllable planning.

We evaluate \textit{JumpStarter} through two complementary studies. A within-subjects user study (N=10) shows that plans produced with \textit{JumpStarter} were rated substantially higher in quality than those produced with \texttt{GPT-4o} via the ChatGPT interface, with significant gains in draft quality, exploration, confidence, and workload. A large-scale simulation study on a synthesized human-grounded benchmark further shows that \textit{JumpStarter} consistently outperforms ChatGPT baselines, planning and memory-agent baselines, and workflow ablations. Together, these findings suggest that effective human-AI planning depends not simply on providing more context, but on attaching the right context to the right subtask at the right time.

Overall, our contributions are three-fold:
\begin{itemize}
    \item We introduce \textit{task-structured context curation}, a framework for complex human-AI planning that combines hierarchical task decomposition with three core mechanisms: context elicitation, selection, and reuse.
    \item We develop \textit{JumpStarter}, an interactive human-AI planning system that helps users initiate and manage complex personal goals by identifying when and how to decompose tasks, while generating personalized drafts grounded in curated task-relevant context.
    \item We empirically validate the approach through human-centered and automatic evaluations, demonstrating clear benefits of task-structured context curation over existing approaches and pointing to a promising direction for personalized, goal-oriented LLM-based systems.
\end{itemize}

\section{Related Work}

\paragraph{LLM Planning Agents and Task Decomposition}
A growing body of work studies how LLMs support goal-directed planning. Benchmarks and systems for travel, scheduling, web navigation, and everyday tasks show that decomposing open-ended goals into intermediate actions or constraints is often what makes them tractable~\cite{lal2024tailoring,xie2024travelplanner,zheng2024natural}. ADaPT~\cite{prasad2024adapt} recursively decomposes tasks and revises plans after execution failures, while autonomous frameworks such as AutoGPT~\cite{Significant_Gravitas_AutoGPT} chain model-generated actions without user involvement. Because user goals are often underspecified, Ask-before-Plan~\cite{zhang2024ask} and proactive dialogue systems~\cite{malaviya2024contextualized,deng2023survey,zhang2023groundialog} study when an agent should ask clarifying questions, and Cocoa~\cite{feng2026cocoa} examines how humans and agents coordinate over shared plans rather than handing off a finished one. \textit{JumpStarter} builds on these mechanisms (decomposition, clarification, and mixed-initiative coordination) but binds each to a user-facing task structure, so that they operate locally at the level of individual subtasks rather than over a single, flat interaction history.

\paragraph{Interactive Context Curation for LLM Workflows}
Memory-augmented agents maintain context across long interactions by storing prior inputs, intermediate results, or observations and retrieving them later; RAISE~\cite{liu2401llm}, for example, keeps short- and long-term memory, while the Model Context Protocol (MCP)~\cite{anthropic2024mcp} standardizes how models access external tools and application state. In HCI, task-centric information management has long shown that structured representations help users externalize goals, manage project materials, and resume interrupted work~\cite{kerne2014using,personal_project_planner,umea,towel,programmer_context}, and ExploreLLM~\cite{explorellm} uses a structured interface to help users decompose goals and specify preferences. \textit{JumpStarter} extends this direction by using the task structure as the organizing unit for eliciting, selecting, and reusing context during personalized planning. Appendix~\ref{app:extended-related} further relates it to automated planning and to task-management and workflow systems.

\section{\textit{JumpStarter} System}
\label{sec:system}

We take a human-centered design approach to developing \textit{JumpStarter}. Rather than treating complex planning as a one-shot generation problem, we design the system around how users progressively clarify goals, surface constraints, make decisions, and produce intermediate artifacts (i.e. answer drafts). This motivates \textit{task-structured context curation}: a workflow that organizes user context around the evolving structure of a complex planning goal. Instead of relying on a flat chat history or a single long-context prompt, \textit{JumpStarter} decomposes a goal into localized subtasks, scopes relevant context to each subtask, and preserves intermediate answer drafts as reusable planning artifacts.

In this section, we first describe the design principles behind \textit{task-structured context curation} and a pilot study that validates these principles through controlled ablations. We then describe how \textit{JumpStarter} implements the workflow through task decomposition and context curation. A full interface walkthrough is provided in Appendix~\ref{sec:sys_walk}.




\subsection{Design Principles}
\label{sec:design-principles}

We build \textit{JumpStarter} around the following four design principles.

\paragraph{DP1: Context Reuse - Preserve intermediate artifacts as reusable context.}
Complex planning often proceeds through partial artifacts, such as schedules, checklists, drafts, comparison tables, or decision notes. Prior work on distributed and external cognition shows that artifacts can serve as external representations that help people organize, inspect, and continue cognitive work across time~\cite{hutchins2000distributed}. \textit{JumpStarter} therefore treats intermediate answer drafts as reusable planning artifacts rather than isolated model responses.

\paragraph{DP2: Context Selection - Scope context to the current subtask.}
Different subtasks require different information. Although long-context LLMs can accept large amounts of input, prior work shows that they do not always use long contexts robustly and can fail to retrieve relevant information when it is buried among irrelevant content~\citep{liu2024lost}. \textit{JumpStarter} therefore selects task-relevant context before each generation step, scoping the model input to the current planning decision.

\paragraph{DP3: Context Elicitation - Elicit missing context when needed.}
Users often do not know what information is needed upfront, especially when goals are underspecified. Mixed-initiative systems have long emphasized that intelligent systems should reason about uncertainty and ask for information when user intent or task requirements are unclear~\cite{horvitz1999principles}. Recent work on proactive planning agents similarly shows the importance of asking clarifying questions before generating plans under ambiguous user instructions~\cite{zhang2024ask}. \textit{JumpStarter} therefore elicits context incrementally, asking targeted questions when missing information is needed for a specific planning decision.

\subsection{Pilot Testing and Design Validation}
\label{sec:pilot-design}

To validate whether these principles provide complementary benefits, we conducted a controlled within-subjects pilot study with six expert evaluators.\footnote{Refer to Appendix \ref{sec:app_tech_eval} for participant backgrounds and the detailed pilot procedure.} The study compared three variants of the same system: (1) \textit{context reuse only}, where previously generated drafts could be reused across subtasks; (2) \textit{context selection and reuse}, where the system additionally selected task-relevant context before generating each draft; and (3) the full \textit{context elicitation, selection, and reuse} pipeline, where the system also asked targeted questions to gather missing context. Apart from these ablated mechanisms, all participants used the same interface, task structure, and model backend (i.e. \texttt{GPT-4o}).

\begin{figure}[t]
    \centering
    \includegraphics[width=0.45\textwidth]{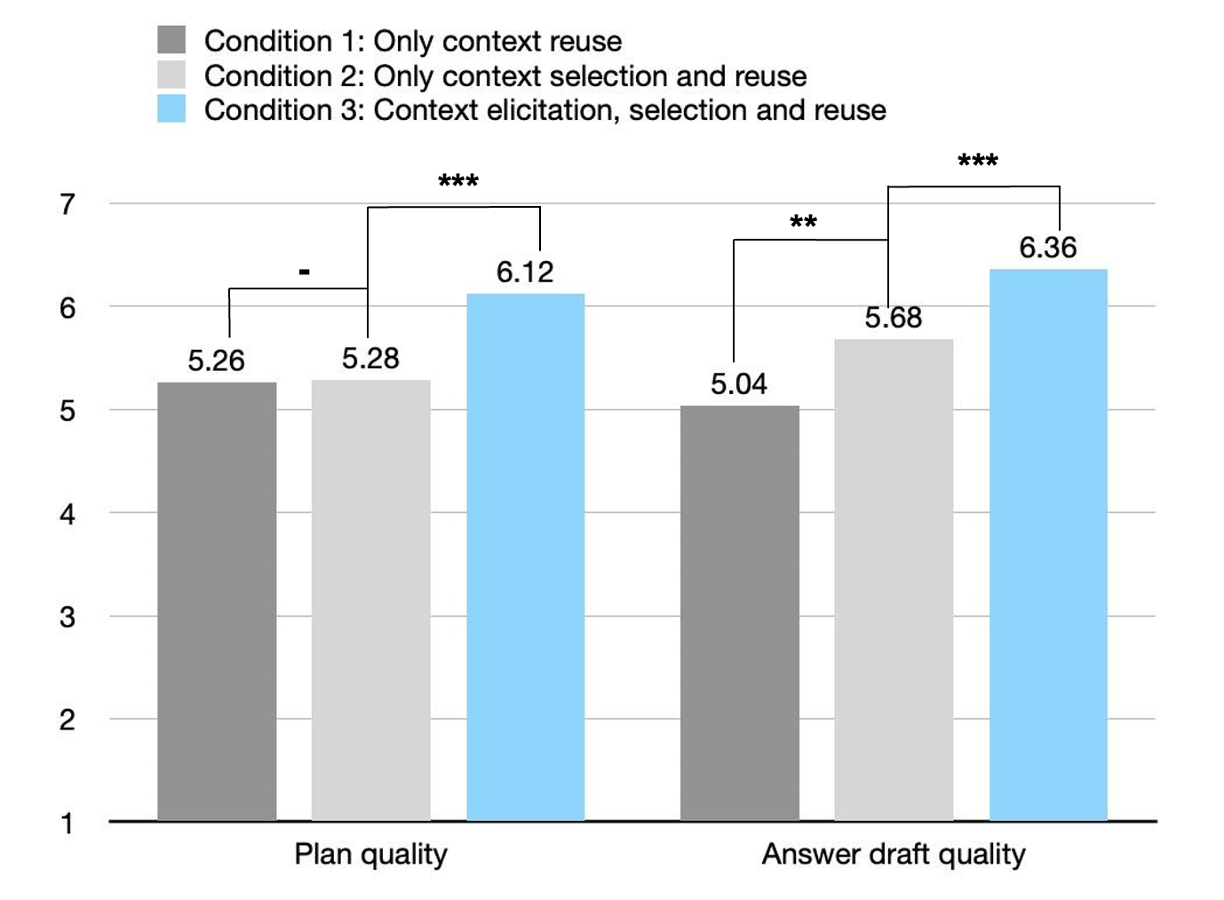}
    \caption{Expert evaluation of plan and answer draft quality across three conditions. Our task-structured context curation method (context elicitation, selection, and reuse) significantly outperforms both context reuse only and context selection and reuse only. Improvements are statistically significant (***p < 0.001, **p < 0.01).} 
\label{fig:tech_eval_results}
\end{figure}

The pilot showed that the full context-curation pipeline produced higher-quality action plans and answer drafts than the ablated variants (see Figure~\ref{fig:tech_eval_results}). These results support the design rationale behind \textit{JumpStarter}: reuse preserves intermediate progress, selection focuses each generation step on relevant information, and elicitation supplies missing information that cannot be recovered from prior context alone. We use these validated mechanisms as the basis for the final system implementation.

\subsection{Task-Structured Context Curation}
\label{sec:jumpstarter}

\textit{JumpStarter} implements \textit{task-structured context curation} as an interactive LLM-based planning workflow. The system uses LLMs in two main ways: \textit{task decomposition}, which organizes a complex goal into actionable subtasks, and \textit{context curation}, which elicits, selects, and reuses information to support each subtask. Together, these operations allow \textit{JumpStarter} to attach context and intermediate artifacts to specific planning steps rather than relying on a flat chat history. Full prompts are provided in Appendix~\ref{sec:prompts}.

\subsubsection{Task Decomposition}
\paragraph{Subtask Generation} Task decomposition provides the structural scaffold for planning. When a user enters a goal or chooses to break down an existing task, \textit{JumpStarter} prompts \texttt{GPT-4o} to generate a set of actionable subtasks. Each subtask includes a name, description, and estimated completion time, helping users understand the scope and effort required. To reduce redundancy, the prompt includes the current task tree and instructs the model to generate subtasks that complement the existing structure (see Figure \ref{app:p_sg}).

\paragraph{Subtask Detection} Before generating a draft for a selected task, \textit{JumpStarter} performs \textit{subtask detection}: it evaluates whether the task is sufficiently detailed and actionable. If the task is too broad, the system recommends further decomposition; otherwise, it proceeds to context curation and draft generation. We implement this feature with \texttt{GPT-4o} using Chain-of-Thought prompting \cite{wei2022chain} with few-shot examples and the task's tree level as additional input. We report the evaluation details for subtask detection in Section~\ref{sec:sub_detect}.

\paragraph{Task Forking} Some tasks require parallel rather than sequential decomposition. 
Consider a user preparing PhD applications: they may need to contact multiple professors, compare multiple institutions, or invite multiple recommenders.
For these cases, \textit{JumpStarter} supports \textit{task forking}, where we prompt \texttt{GPT-4o} to determine whether a task should be decomposed into entity-specific subtasks (see Figure \ref{app:p_forking}). When forking is appropriate, the system first identifies the context relevant to the entities involved, then generates parallel subtasks grounded in those selected context elements. This helps produce subtasks that are independent and non-redundant.

\subsubsection{Context Curation}
Context curation operationalizes the three mechanisms of task-structured context curation: elicitation, selection, and reuse. 
\paragraph{Context Elicitation} At goal initialization, \textit{JumpStarter} prompts \texttt{GPT-4o} to identify information that would help personalize the plan, including constraints, preferences, deadlines, resources, prior progress, and relevant documents (see Figure \ref{app:p_ce_root}). The system presents these as elicitation questions or document suggestions, and the user's responses are stored as global context for the planning session.

\paragraph{Context Selection}For each subtask, \textit{JumpStarter} performs task-local context selection before draft generation. The system constructs a candidate context pool from global context, local user responses, saved drafts, and related subtasks. Given the selected task's title and description, \texttt{GPT-4o} selects the context entries most relevant to the current subtask.\footnote{See prompts in Appendix \ref{app:prompt_cc}} These selected context keys are shown to the user for review, allowing the user to remove irrelevant items or add missing information before generation. This step scopes generation to the current planning decision instead of indiscriminately passing all available context to the model.

\paragraph{Context Reuse and Answer Draft Generation} After context selection, \textit{JumpStarter} prompts \texttt{GPT-4o} with the user's goal, the selected subtask, global context, and task-local selected context to generate an answer draft. The draft is designed to be a concrete intermediate artifact, such as a checklist, schedule, or message. When the user approves a draft, the system stores it as a reusable local context. Later subtasks can select and reuse these saved drafts, allowing \textit{JumpStarter} to carry forward refined intermediate artifacts throughout the planning workflow. When users want to improve a draft but are unsure what additional information would help, \textit{JumpStarter} generates targeted follow-up questions for the current subtask (see Figure \ref{app:p_ce_drafting}). The user's responses are added to the local context, and the system regenerates the draft using the updated context.

\section{Evaluation of \textit{JumpStarter}}

We evaluate \textit{JumpStarter} through two complementary studies. A within-subjects user study compares it with ChatGPT on participants' own planning goals, measuring planning experience, output quality, confidence, and workload. A large-scale simulation study on a human-grounded benchmark built from the user-study traces then tests whether these benefits hold against strong baselines and ablations.


\subsection{User Study}
\label{sec:user-study}

We conducted a within-subjects study with ten participants comparing \textit{JumpStarter} against ChatGPT.\footnote{We used the ChatGPT interface with \texttt{GPT-4o} specified in the user study.} We selected ChatGPT as the baseline because it is familiar to users and supports contextual interaction across a session, making it a realistic comparison for goal-oriented planning.

Each participant used both systems to explore a
personal goal they intended to pursue in the near
future. After interacting with each system, participants completed questionnaires assessing task load and satisfaction with outcomes. We also conducted semi-structured interviews to gather qualitative insights into their experience with each system.

\subsubsection{Participants and Procedure}

\begin{table}[t]
\centering
\begin{adjustbox}{max width=\linewidth}
    \begin{tabular}{| l | l | l | }
    \hline 
         \textbf{Participant} &\textbf{Personal goal} &\textbf{Goal type}\\ 
         \hline
         \textbf{P1}&  Start a side job &  Career \\ 
         \hline 
          \textbf{P2}& Organize a weekly game night & Life   \\ 
         \hline 
        \textbf{P3}&  Land a job offer &  Career  \\ 
        \hline
         \textbf{P4}& Prepare for the LSAT  & Academia   \\ 
         \hline 
         \textbf{P5}& Manage social media accounts & Creativity   \\ 
         \hline 
         \textbf{P6}&  Move to a new apartment  & Life  \\ 
         \hline 
         \textbf{P7}&  Create a portfolio website & Creativity   \\ 
         \hline 
         \textbf{P8}&  Prepare to deliver a tutorial   & Academia \\ 
         \hline 
         \textbf{P9}&  Start a YouTube channel& Creativity \\ 
         \hline 
        \textbf{P10}&  Organize a family reunion& Life \\ 
         \hline 

    \end{tabular}
    \end{adjustbox}
    \caption{Overview of personal goals picked by participants in the user study.}
    \label{tab:user_picking_goals}
\end{table}

We recruited ten participants (average age=23.8;
six female, four male) through a university mailing
list and word of mouth. All reported being familiar
or very familiar with ChatGPT. Before the study,
each participant selected a personal goal to pursue
within the next six months (see Table \ref{tab:user_picking_goals}). At the
start of each session, participants were introduced
to the concepts of action plans and answer drafts
through examples, then used both ChatGPT and JumpStarter to plan and generate answer drafts for as many subtasks as possible. The system order was randomized and counterbalanced to mitigate
order effects. Each task was limited to 25 minutes, and each study session lasted approximately 1.5 hours. All participants were compensated \$20 per hour. Appendix~\ref{sec:details_user_study} discusses the scope of the study.

\subsubsection{Results and Findings}

\begin{figure}[t]
    \centering
    \begin{subfigure}[b]{0.45\textwidth}
        \centering
        \includegraphics[width=\textwidth]{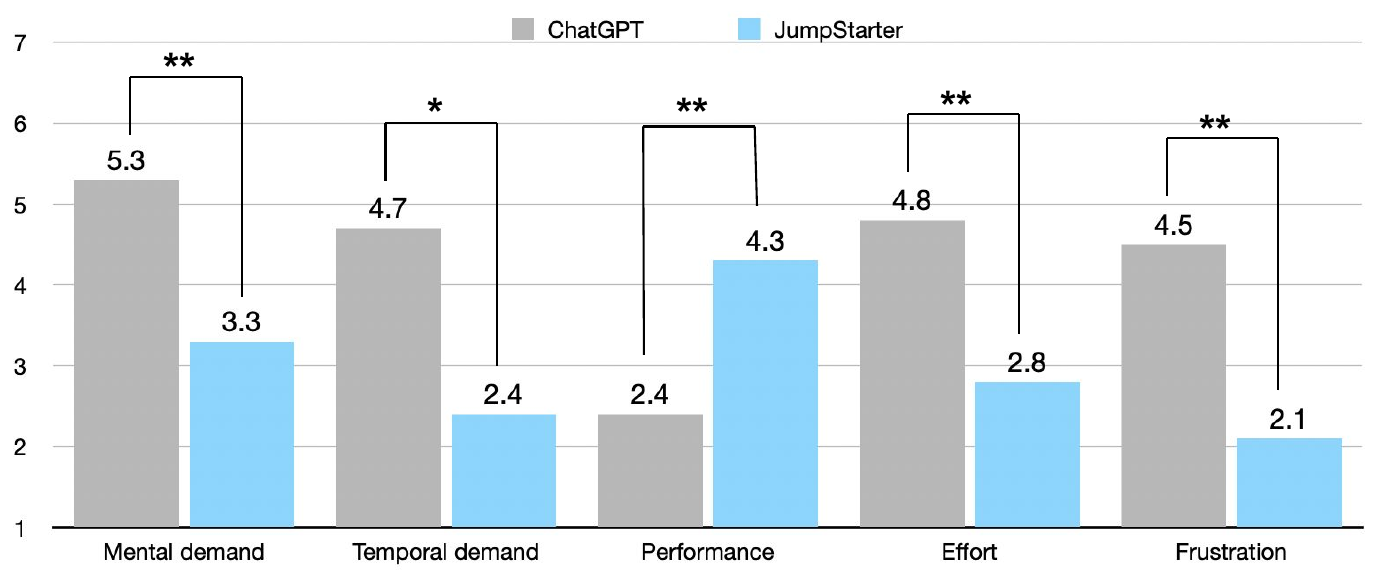}
        \caption{NASA Task Load Index Comparison}
        \label{fig:nasa}
    \end{subfigure}

    \vspace{0.2cm}

    \begin{subfigure}[b]{0.4\textwidth}
        \centering
        \includegraphics[width=\textwidth]{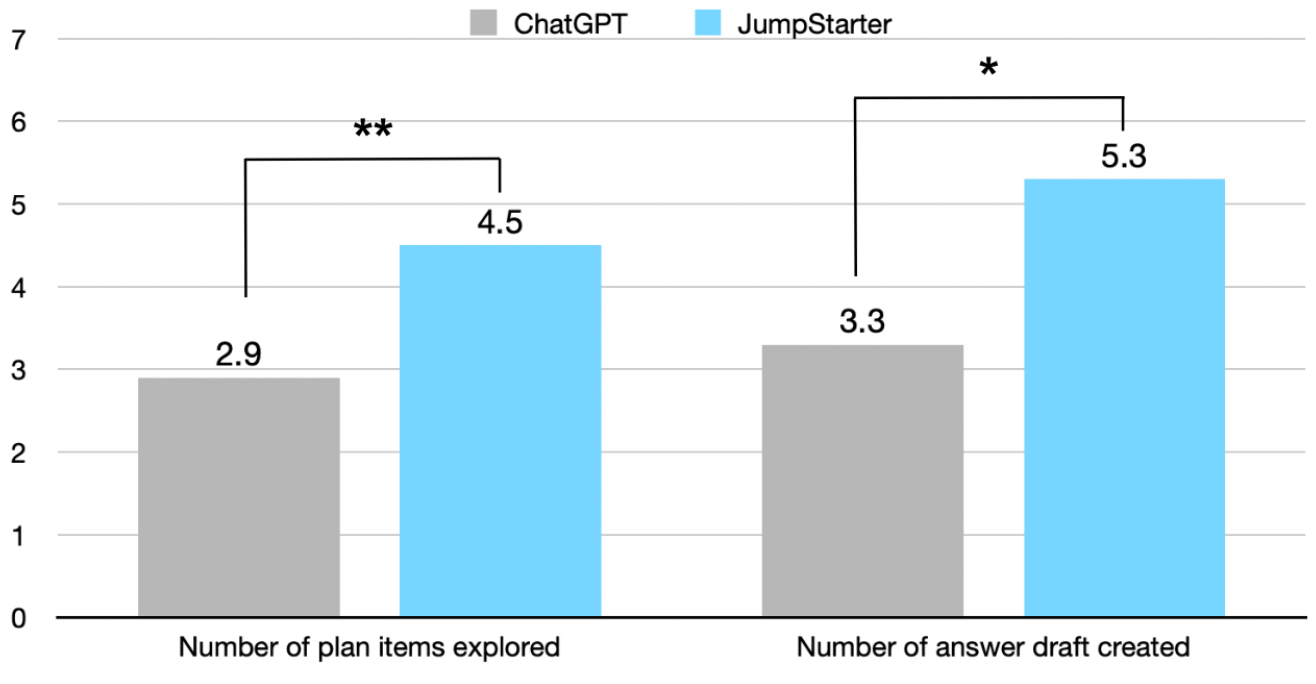}
        \caption{Exploration Efficiency Comparison}
        \label{fig:exploration}
    \end{subfigure}

    \vspace{0.2cm}

    \begin{subfigure}[b]{0.4\textwidth}
        \centering
        \includegraphics[width=\textwidth]{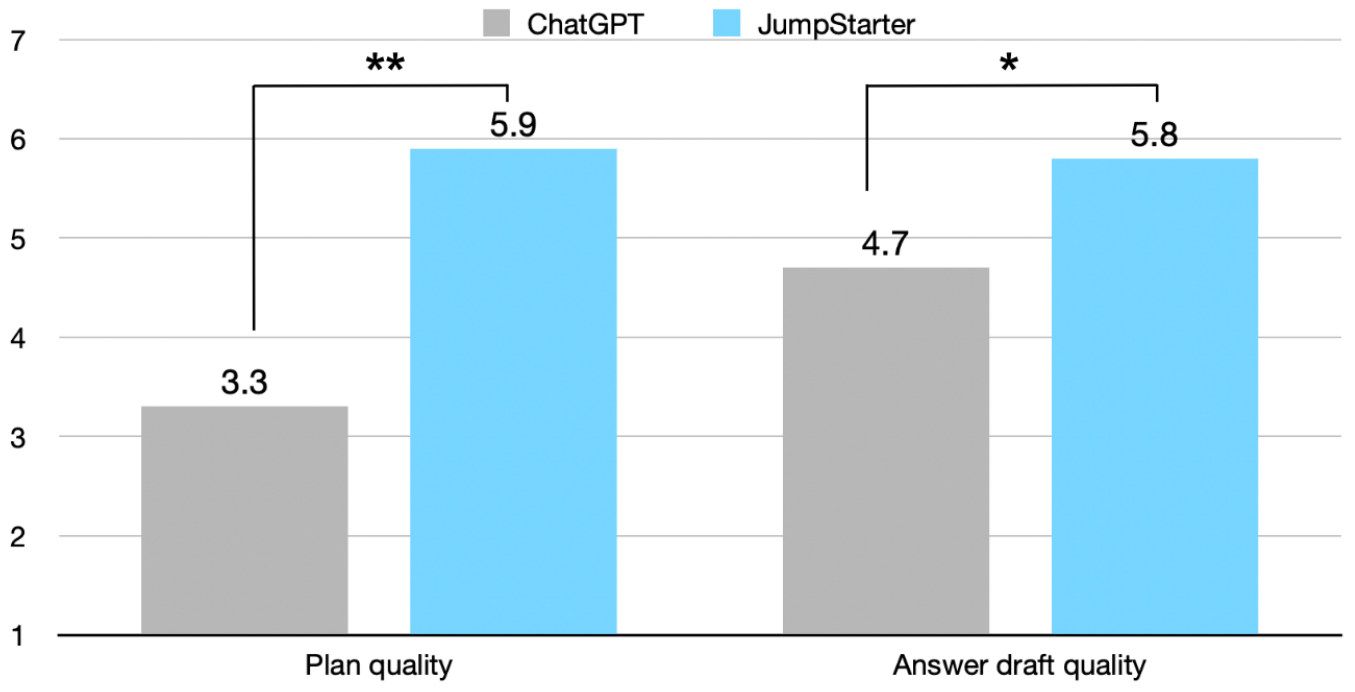}
        \caption{Result Satisfaction Comparison}
        \label{fig:satisfaction}
    \end{subfigure}

    \caption{User study results comparison between using ChatGPT and using \textit{JumpStarter}. The statistical test results comparing JumpStarter with ChatGPT, where the p-values ($*$: $p<.050$, $**$: $p<.010$, $***$: $p<.001$) are reported. } 
    \label{fig:user_study_results}
\end{figure}

We collected 1--7 scale questionnaire ratings on task load, outcome satisfaction, and confidence, and recorded the number of plan items explored and answer drafts generated by each participant. We further analyzed semi-structured interviews using thematic analysis~\cite{braun2006using}. Figure~\ref{fig:user_study_results} summarizes the main results; additional details are provided in Appendix~\ref{sec:details_user_study}.

\paragraph{\textit{JumpStarter} Reduces Users' Task Load}
As shown in Figure~\ref{fig:nasa}, participants reported significantly lower workload with \textit{JumpStarter} than with ChatGPT across mental demand ($p=.005$), temporal demand ($p=.012$), performance ($p=.007$), effort ($p=.007$), and frustration ($p=.009$). Participants attributed this reduction to \textit{JumpStarter}'s task structure and context-aware guidance, which reduced the need to manually maintain planning state and decide what information to provide next. For example, P2 noted that ChatGPT required them to keep the structure in their head, whereas \textit{JumpStarter} provided a structure they could follow.

\paragraph{\textit{JumpStarter} Improves Exploration Efficiency.}
Participants explored more plan items with \textit{JumpStarter} than with ChatGPT (4.5 vs. 2.9), and generated more answer drafts (5.3 vs. 3.3). This suggests that \textit{JumpStarter}'s task-structured workflow helped users make broader progress across their goals. In interviews, participants explained that ChatGPT's linear format often led them to over-iterate on a single thread, while \textit{JumpStarter} made it easier to move across subtasks while preserving relevant context and prior drafts.

\paragraph{\textit{JumpStarter} Improves Perceived Outcome Quality.}
Participants rated plans produced with \textit{JumpStarter} as higher quality than those produced with ChatGPT (5.9 vs. 3.3), and also rated answer drafts higher (5.8 vs. 4.7). Participants attributed this improvement to targeted elicitation and task-relevant context use. For example, P9 noted that \textit{JumpStarter} incorporated their prior video-editing experience into the YouTube-channel plan, while ChatGPT focused on editing advice they did not need.

\subsection{Human-Grounded Automatic Simulation}
\label{sec:simulation}

We complement the user study with an automatic simulation study to examines whether the user-study findings generalize across a broader set of goals and personas, while systematically comparing \textit{JumpStarter} against relevant competitive baselines and ablations.

\subsubsection{Dataset Construction}

We construct a human-grounded simulation benchmark from realistic planning goals and persona-based simulated users. 
For goals, we seed the benchmark with 13 logged study goals (10 from the user study, 3 from the pilot) and synthesize 47 additional goals with \texttt{GPT-5.5}, yielding 60 personalized planning tasks stratified across five domains (see Table~\ref{tab:appendix-domain-quotas}). Each synthesized candidate must pass an LLM-based rubric filter that enforces realism, personalization, decomposability, and tangible outputs, as well as manual inspection by the leading authors.
For users, we extract 10 base personas from user-study traces and create 10 controlled variants, yielding 20 personas. Each persona specifies interaction-relevant attributes such as expertise, context richness, decision-making style, communication style, and information that should only be revealed when explicitly asked. We provide details on goal synthesis, persona construction, and validation in Appendices~\ref{app:details_dataset_con}.

Crossing 60 goals with 20 personas yields 1,200 possible profiles. We reserve 10 profiles for development. For the main evaluation, we sample 300 of the remaining profiles, balanced so that each goal appears five times and each persona fifteen times.\footnote{We use 300 profiles due to budget constraints. We release all 1,200 profiles to support larger-scale evaluation.}
Because LLM-based user simulators can be overly cooperative and may not reliably reflect diverse human behaviors \cite{seshadri2026lost}, we calibrate simulated-user decisions using real user study logs, including task-tree acceptance, draft saving, node completion, and revision behavior. Details of the calibration process and example simulation profiles are provided in Appendix~\ref{app:sim_users}.

\subsubsection{Baseline Methods}

We compare \textit{JumpStarter} against three groups of baselines. 
\textbf{\textit{ChatGPT baselines}} use \texttt{GPT-4o} in a multi-turn planning setting that mirrors the user-study ChatGPT condition: \textit{vanilla ChatGPT}, \textit{ChatGPT + elicited context}, and \textit{ChatGPT + structured summary}. These baselines test whether improvements come from additional user information or a better-organized prompt. We further add \textit{single-turn decomposition}, which receives the same elicited context and is explicitly prompted to decompose the goal and draft every subtask in one turn, testing whether \texttt{GPT-4o}'s own decomposition can replace an external task structure.
\textbf{\textit{Planning and memory-agent baselines}} represent alternative agentic approaches to complex planning: \textit{ADaPT} \citep{prasad2024adapt} for recursive task decomposition, \textit{Ask-Before-Plan} \citep{zhang2024ask} for clarification-first planning, an unstructured memory-RAG baseline inspired by memory-augmented agents such as MemGPT \citep{packer2023memgpt}, and an \textit{integrated agentic planner} that elicits global context, retrieves from the user's materials, queries a simulated browse tool, and selects context for each subtask, but does not reuse drafts across subtasks. 
These baselines test whether decomposition, clarification, retrieval, or their combination can match task-structured context curation. \textbf{\textit{JumpStarter workflow ablations}} isolate the contribution of \textit{JumpStarter}'s mechanisms: \textit{random context selection}, \textit{no context selection}, \textit{no context reuse}, and \textit{no elicitation}. These conditions test whether performance depends on task-relevant context selection, intermediate artifact reuse, proactive elicitation, or merely exposing the model to more context. All methods are evaluated on the same 300 simulation profiles, with \texttt{GPT-4o} used for workflow execution and simulated-user responses. Detailed baseline descriptions are provided in Appendix~\ref{app:details_baseline}.

\subsubsection{Metrics and Judging}

We use \textit{GPT-5.5} as a blinded judge to evaluate the final outputs from each condition. To reduce presentation bias, we remove method names before judging. 
The judge scores each output from 1 to 7 on the main outcome dimensions from the human study: plan quality, answer draft quality, confidence, personalization, and reduced burden. Our primary metric is a weighted composite of these dimensions. The full rubric and judging prompt are provided in Appendix~\ref{app:llm-as-judge}.


Before applying the judge to the full test set, we validate it on 10 matched within-subject JumpStarter--ChatGPT pairs from the user study, yielding 20 de-identified system outputs. The judge achieves a Spearman correlation of 0.779 with human ratings, suggesting alignment with the study construct. The full validation procedure is provided in Appendix~\ref{app:llm-as-judge}.

\subsubsection{Results and Findings}
\label{sec:simulation-results}

We report \textit{JumpStarter-Shallow} as the primary condition because it best matches how participants used the system in the real user study: a lightweight task structure with at most two decompositions beyond the initial subtask layer (see Figure \ref{fig:teaser}). We also evaluate \textit{JumpStarter-Recursive}, which uses the same context-curation mechanisms but allows subtasks to be decomposed further. Table~\ref{tab:auto-results} summarizes the results. \textit{JumpStarter-Shallow} achieves the highest overall quality score (4.27), and positive deltas indicate matched improvements over each baseline.

\begin{table}[t]
\centering
\setlength{\tabcolsep}{1pt}
\begin{tabular}{lcc}
\toprule
\textbf{Method} & \textbf{Score} & \textbf{$\Delta$} \\
\midrule
\multicolumn{3}{l}{\textit{\textbf{ChatGPT baselines}}} \\
ChatGPT vanilla & 3.75 & \textcolor{posgreen}{\textbf{+0.52}} \\
+ elicited context & 3.70 & \textcolor{posgreen}{\textbf{+0.57}} \\
+ structured summary & 3.72 & \textcolor{posgreen}{\textbf{+0.55}} \\
Single-turn decomposition & 4.05 & \textcolor{posgreen}{\textbf{+0.22}} \\
\midrule
\multicolumn{3}{l}{\textit{\textbf{Planning/memory baselines}}} \\
ADaPT \cite{prasad2024adapt} & 4.05 & \textcolor{posgreen}{\textbf{+0.22}} \\
Ask-before-plan \cite{zhang2024ask} & 3.22 & \textcolor{posgreen}{\textbf{+1.05}} \\
Memory-RAG \cite{packer2023memgpt} & 3.29 & \textcolor{posgreen}{\textbf{+0.98}} \\
Integrated agentic planner & 3.94 & \textcolor{posgreen}{\textbf{+0.33}} \\
\midrule
\multicolumn{3}{l}{\textit{\textbf{JumpStarter ablations}}} \\
Random context selection & 4.14 & \textcolor{posgreen}{\textbf{+0.13}} \\
No context selection & 4.08 & \textcolor{posgreen}{\textbf{+0.19}} \\
No context reuse & 4.09 & \textcolor{posgreen}{\textbf{+0.18}} \\
No context elicitation & 4.19 & \textcolor{posgreen}{\textbf{+0.08}} \\
\midrule
\textit{JumpStarter--Recursive} & 4.20 & \textcolor{posgreen}{\textbf{+0.07}} \\
{\textit{JumpStarter--Shallow}} & \textbf{4.27} & -- \\
\bottomrule
\end{tabular}
\caption{Automatic simulation results on 300 held-out persona-goal profiles. Scores are blinded judge study-quality composites. $\Delta$ reports the matched difference against \textit{JumpStarter-Shallow}. Green deltas indicate statistically significant improvements where the 95\% confidence interval of the paired difference excludes zero. Full confidence intervals are reported in Appendix~\ref{app-tab:auto-results}.}
\label{tab:auto-results}
\end{table}

\paragraph{JumpStarter outperforms ChatGPT baselines.}
\textit{JumpStarter-Shallow} substantially outperforms vanilla ChatGPT (+0.52), ChatGPT with elicited context (+0.57), and ChatGPT with a structured context summary (+0.55). Since the latter two baselines receive the same elicited user information, these results suggest that the gains come from organizing context through the task-structured workflow, not merely from providing more user information to the model.




\paragraph{Context curation matters.}

Ablation results confirm that each context-curation mechanism contributes to performance. As shown in Figure~\ref{fig:quality_context_eff}, \textit{JumpStarter-Shallow} achieves the best quality-efficiency tradeoff: it outperforms the no-selection baseline while using fewer input tokens on average ($19,158$ vs. $26,955$). A GPT-based relevance labeler further shows that \textit{JumpStarter-Shallow} selects substantially more relevant context than the no-selection baseline ($0.72$ vs. $0.22$ relevance score; Appendix~\ref{app:ctxt_label}). Removing context reuse lowers the score from $4.27$ to $4.09$, randomizing context selection lowers it to $4.14$, and removing elicitation lowers it to $4.19$. Together, these results show that performance depends on selecting task-relevant context, preserving intermediate artifacts, and eliciting missing information when needed, rather than increasing context volume alone.

\paragraph{Generic planning and memory agents do not replace workflow structure.}
\textit{JumpStarter-Shallow} outperforms ADaPT (+0.22), Ask-before-plan (+1.05), and memory-RAG (+0.98). It also outperforms the integrated agentic planner (+0.33), which elicits, retrieves, and selects context dynamically; the agent's largest deficits are in personalization (+0.63) and in context curation and reuse (+0.58).\footnote{Mean difference (\textit{JumpStarter-Shallow} minus agent) in the judge's 1--7 scores on each criterion; see the rubric in Appendix~\ref{app:llm-as-judge}.}
Appendix~\ref{app:integrated-agent} analyzes this comparison in detail. These results suggest that decomposition, clarification, or memory retrieval alone is insufficient for personalized planning; the key advantage is using task structure to transform context into local artifacts and reuse them across the workflow.


\begin{figure}[t]
    \centering
    \includegraphics[width=0.45\textwidth]{figures/quality_context_efficiency.pdf}
    \caption{JumpStarter ablation results comparing study quality against input context-token usage. Each point reports the mean quality score and average input context tokens per task over $300$ persona-goal profiles. \textit{JumpStarter-Shallow} offers the strongest quality-efficiency tradeoff, achieving the highest quality score while using fewer context tokens than most ablations.\protect\footnotemark}
    \label{fig:quality_context_eff}
\end{figure}
\footnotetext{We use \texttt{tiktoken} to compute token counts.}

\paragraph{Lightweight task structure is more effective than deeper recursion.}
\textit{JumpStarter-Recursive} is competitive (4.20), but \textit{JumpStarter-Shallow} performs better (+0.07). This aligns with our real-use traces, where participants generally used \textit{JumpStarter} as a shallow planning workspace rather than a deeply recursive planner. The strongest configuration provides enough structure to support context selection and reuse without over-decomposing the task.

\subsubsection{Failure Case Analysis}
\label{sec:failure-cases}

We compare \textit{JumpStarter-Shallow} with its strongest one-shot competitor, single-turn decomposition, on each profile. \textit{JumpStarter-Shallow} loses on 31\% of the profiles, and more often in a few settings (Appendix Table~\ref{tab:failure-breakdown}).


\paragraph{Gains shrink on simple, single-artifact goals.}
On simple everyday-administration goals, such as decluttering a home office, the relative gain in study quality falls from 5.4\% to 2.6\% and the loss rate rises to 38\%. Losses are also more frequent for goals spanning weeks rather than months (36\% vs.\ 29\%), for career and education goals (38\%), and for users who volunteer little context (40\%). \textit{JumpStarter-Recursive} gains 6.0\% on complex goals but only 1.3\% on simple ones, which is not significant.

\paragraph{Failures stem from fragmented outputs, not poor grounding.}
On the profiles it loses, \textit{JumpStarter-Shallow} still scores 22\% higher on context curation and reuse and matches the baseline on personalization. However, it scores 13--18\% lower on plan quality, tangible results, confidence, and user burden: its output is about three times longer and split across subtask drafts. For decluttering a home office, it produced four overlapping drafts that left undecided details as placeholders, and it scored 22\% lower than the baseline's single checklist-based plan. Deeper recursion adds inconsistency: \textit{JumpStarter-Recursive} loses to \textit{JumpStarter-Shallow} on 61\% of simple goals and scores 23\% lower on avoiding contradictions. Task-structured curation therefore pays off on complex, multi-deliverable goals, while simple goals are better served by a single consolidated draft.

\section{Conclusion}


We introduced \textit{task-structured context curation}, a framework for human-AI planning that organizes user context around the evolving structure of a complex goal. Implemented in \textit{JumpStarter}, this approach decomposes goals into actionable subtasks while eliciting missing information, selecting task-relevant context, and reusing intermediate artifacts. A within-subjects user study shows that \textit{JumpStarter} improves plan and draft quality, exploration, confidence, and workload compared to ChatGPT. A human-grounded simulation study further shows that these gains persist across broader goals and personas, and cannot be explained by stronger prompting, additional context, retrieval, clarification-first planning, or deeper recursion alone. Together, our findings suggest that effective human-AI planning depends on attaching the right context to the right subtask at the right time.

\section{Limitations}
\label{sec:limitations}

\textit{JumpStarter} helps users reason about \textit{how} to pursue personal goals by making them more concrete, structured, and actionable. 
However, it does not directly address the \textit{why} dimension of goal pursuit, such as motivation, self-regulation, or emotional challenges. 
For example, one participant reported lower confidence after realizing how much work their goal required. 
While motivational support could help, prior work cautions that LLM-based emotional support raises safety and ethical concerns~\cite{joseph}. 
Future work should explore how planning systems can provide motivational scaffolding responsibly.

Our evaluation combines a real user study with a larger automatic benchmark, but simulated users and LLM-based judges cannot fully replace human behavior and evaluation. 
Simulated users may be more cooperative or consistent than real users, and LLM judges may introduce model-specific biases. 
Future work should validate these findings through larger human studies or longitudinal deployments.

\textit{JumpStarter} focuses on cognitive and knowledge-work goals, including academia, career development, creativity, and everyday life. 
It does not yet address physical, behavioral, or spiritual goals, such as losing weight, overcoming shyness, or coming to terms with one's faith~\cite{little1983personal}. 
These goals often require habit formation, accountability, or motivational support~\cite{consolvo2009goal,dis_goaL_setting}, which remain directions for future work.

Finally, \textit{JumpStarter} relies on \texttt{GPT-4o} as its core generation engine. 
Although LLMs can generate useful plans, they remain prone to factual errors and hallucinations~\cite{gpt-4}. 
Future work should integrate retrieval, verification, or specialized search agents to improve factual grounding for goals requiring up-to-date or domain-specific information.

\section{Ethics Statement}

\paragraph{Biases}
\textit{JumpStarter} relies on pre-trained LLMs, which may reflect biases present in their training data and could produce unfair, stereotypical, or otherwise inappropriate suggestions. 
We did not explicitly mitigate model bias in this work. 
To reduce potential harm, our system is designed for human-AI collaboration rather than autonomous decision-making: users remain in control of selecting context, reviewing outputs, and deciding whether to act on generated plans.

\paragraph{Reproducibility}
We use \texttt{GPT-4o} as the backbone of \textit{JumpStarter}. 
Because proprietary LLMs may change over time, exact reproduction of model outputs may be difficult. 
To support reproducibility, we provide the prompts, system design details, and evaluation protocols used in our experiments.

\paragraph{Study Participants}
We recruited participants for the pilot study and the real user study through a university mailing list and word of mouth. 
Participants were compensated at a rate of \$20 per hour. 
To protect privacy and anonymity, we will not release participants' personal information, demographic details, raw interaction logs, or goal-related materials that may contain sensitive personal context. 
Any released examples will be anonymized.






\section*{Acknowledgments}
We thank the members of the Columbia NLP Lab and the Computational Design Lab at Columbia University for helpful discussions of this work. We are grateful to the participants of our user study for their time and insights, and to the anonymous reviewers for their constructive feedback, which helped improve the paper.

\bibliography{custom}

\appendix

\clearpage

\section{Problem Scope and Extended Related Work}
\label{app:extended-related}

\paragraph{Problem Scope}
We focus on a specific class of planning problems: open-ended, long-horizon planning of personal knowledge-work goals, such as preparing a graduate application or organizing a family reunion. Unlike the problems studied in automated planning, they have no formal domain model (no state space, action schema, or goal predicate); the goal is under-specified and must be elicited from the user and refined during the session; the relevant state is the user's evolving knowledge and context rather than a world state; and there is no ground-truth plan, so success is the user's judgment of plan quality. A person stays in the loop, and the output is a set of interdependent deliverables that span weeks or months. We therefore address the phase before formal planning can begin: turning an ill-defined goal and scattered context into a structure a person can act on.

\paragraph{Relation to Automated Planning}
Classical planning assumes a formal domain model: a state space, actions with preconditions and effects, and a goal condition, as in STRIPS~\cite{fikes1971strips} and PDDL~\cite{aeronautiques1998pddl}, over which a planner searches for a valid or optimal action sequence~\cite{nau2004automated}. Hierarchical task network (HTN) planning is the closest analogue to \textit{JumpStarter} because it also decomposes tasks, but it relies on a hand-authored library of decomposition methods over a well-defined domain~\cite{erol1994htn,nau2003shop2}. The problems \textit{JumpStarter} targets arise before this machinery can run: no domain model exists, and the goal itself must be surfaced and structured with the user. \textit{JumpStarter} supports this pre-formalization phase, and its task structure organizes context for an LLM rather than actions for a planner.

\paragraph{Task Management and Workflow Systems}
Organizing work around tasks has a long history. Workflow management systems model and automate business processes~\cite{georgakopoulos1995overview,van2002workflow}, hierarchical task analysis describes the goals and subgoals of human work~\cite{annett1967task,card2018psychology}, personal task managers track to-dos~\cite{bellotti2004studies}, and coordination theory studies dependencies between activities~\cite{malone1994interdisciplinary}. These systems organize tasks for people to execute and coordinate, so the unit they track is an activity. \textit{JumpStarter} uses task structure for a different purpose: each subtask holds elicited, selected, and reused context, and the operations on it exist to give an LLM the right context for that subtask.

\section{\textit{JumpStarter} System Walkthrough}
\label{sec:sys_walk}


To illustrate how users interact with \textit{JumpStarter}, we present a walkthrough using a concrete example.
Consider John, a user aiming to apply for a PhD in NLP. 
Figures in the Section \ref{sec:sys_vis} provide corresponding interface visuals.

\subsection{Goal Input and Global Context Elicitation}
\label{sec:ce_goal}

John begins by entering his goal—\textit{"Apply for a PhD in NLP"}—into a text input box and clicks \textit{Start}.
The system then generates elicitation questions to collect relevant context (e.g., existing documents, preferences).
In John’s case, it asks about potential target schools and recommendation letters.
Since he is uncertain about the former and has not yet obtained the latter, he uploads his CV in response to the first question and clicks \textit{Let’s get started}. 
The elicited context becomes part of the global context used across subsequent tasks.

\subsection{Subtask Generation and Detection}

Using the elicited context, \textit{JumpStarter} generates an initial subtask tree, presenting titles, descriptions, and estimated durations (Figure~\ref{fig1:init_tree}). 
John reviews the list to gain an overview of the plan and decides to explore the subtasks sequentially.

John selects the first subtask, \textit{Identify Potential PhD Programs}.
The system detects that it is insufficiently actionable and prompts John to either decompose the task or proceed with drafting. 
Following the system's recommendation, John selects \textit{Decompose the task}, resulting in a new set of subtasks under the original node (Figure~\ref{fig2:sd_identify}).

\subsection{Answer Draft Creation and Refinement}

John selects the subtask \textit{Research Universities and Programs}, which the system deems actionable. 
It generates an initial answer draft—a list of NLP PhD programs. 
Upon review, John refines the output by adding, \textit{"I want schools in the Midwest of the US."} 
He is satisfied with the revised list and saves it as an answer draft. 
He is also given three refinement options: regenerate, add more context and regenerate, or iterate on the current draft. 
Saved drafts are stored as context and appear as icons on the task tree, marking completed nodes.

\subsection{Task Forking}

John proceeds to the next subtask, \textit{Identify Faculty Members}. 
The system suggests decomposing the task and detects that it requires forking based on the previously identified programs. 
It selects the saved university list as relevant context and asks John to confirm or modify the selection (Figure~\ref{fig2:cs_forking}). 
Upon confirmation, the system forks the task into program-specific subtasks (Figure~\ref{fig2:nodes_forking}), which John begins to explore individually.


John then explores the task \textit{Get Recommendation Letters} and decomposes it as prompted. 
For the subtask \textit{Compile a List of Recommenders}, the system uses his CV to generate a list including Prof. Blake White, Prof. Julian Deng, and Dr. Alice Feng. 
John accepts and saves the draft. He proceeds to \textit{Reach Out to Potential Recommenders}, which is forked into person-specific subtasks. 
For \textit{Reach Out to Prof. Blake White}, the system identifies relevant context—including John’s CV, prior collaborations, and the university list—and generates a personalized email draft (Figure~\ref{fig2:context-selection-drafting}).

\subsection{Context Elicitation for Draft Iteration}

Unsatisfied with the initial email draft, John clicks \textit{Add Context and Regenerate}. 
The system prompts follow-up questions to elicit additional details (e.g., specific projects or papers), which John provides (Figure~\ref{fig2:context-elicitation-drafting}). 
The refined draft incorporates these details, resulting in a more personalized and acceptable version. 


Through structuring the goal as a hierarchy of tasks, \textit{JumpStarter} helps John curate relevant context, enabling the creation of detailed action plans and high-quality answer drafts. \textit{JumpStarter}'s structured context management and iterative refinement enable him to effectively progress toward his goal of applying for a PhD in NLP.


\subsection{Data Representation}

\textit{JumpStarter} represents each piece of context as a key-value pair, where the key denotes the context name and the value specifies its content. 
For example, the pair \textit{“Location preference: Midwest of US”} captures a user’s geographical preference. 
The system maintains two types of context: global and local. 
Global context consists of information elicited from the user immediately after goal specification and is universally applied throughout planning. 
Local context includes answer drafts and any additional information the user provides during interaction.

To organize tasks, \textit{JumpStarter} employs a hierarchical tree structure.
This design mirrors the natural decomposition of complex goals into manageable components, facilitating clear tracking of task dependencies and progression. 
Each user goal serves as the root node, with system-generated subtasks represented as child nodes. 
Each subtask node stores task-specific attributes, including titles, descriptions, estimated durations, and any associated answer drafts.

\subsection{System visuals}
\label{sec:sys_vis}
\begin{figure*}[ht]
\centering
        \includegraphics[width=0.95\textwidth]{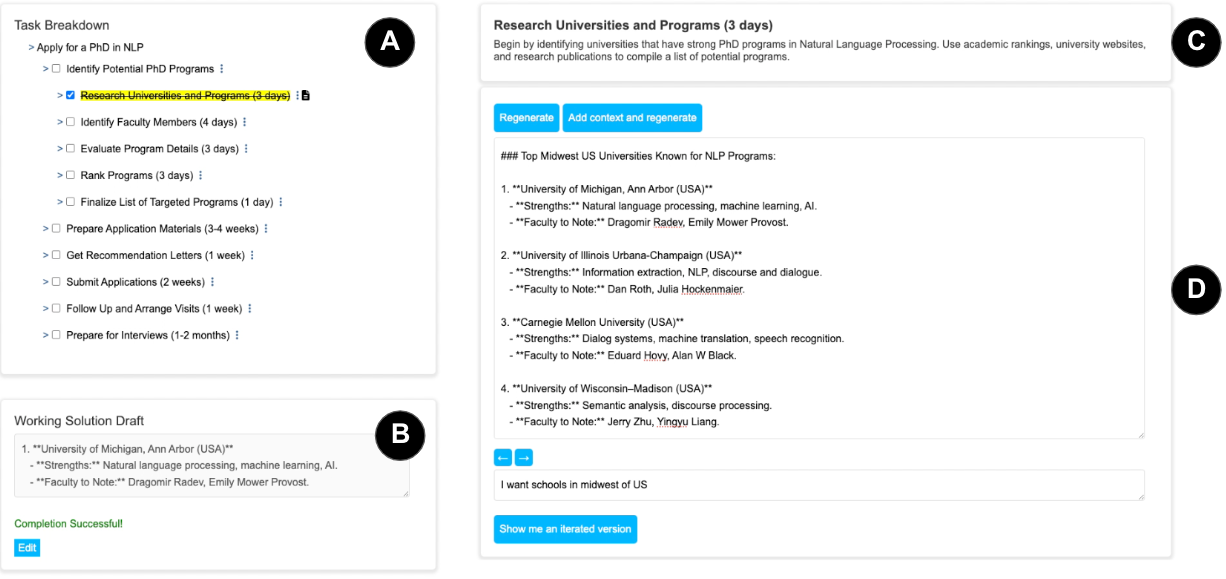}
    \caption{A screenshot of JumpStarter creating plans and answer drafts for the goal \textit{Apply for a PhD in NLP}. (A) Task breakdown is shown as a subtask tree, with the goal being the root node. Subtasks decomposed from the same parent node are shown on the same level. (B) Saving the answer draft. (C) Detailed descriptions of the selected subtask are shown. (D) The answer draft is generated, considering the specification from the user -- "\textit{I want schools in midwest of US}". Users have three options to improve the draft: regenerate, add context and regenerate, and iterate based on users' new specifications. }
    \label{fig:screen_all}
\end{figure*}

\begin{figure}[ht]
\centering
        \includegraphics[width=\linewidth]{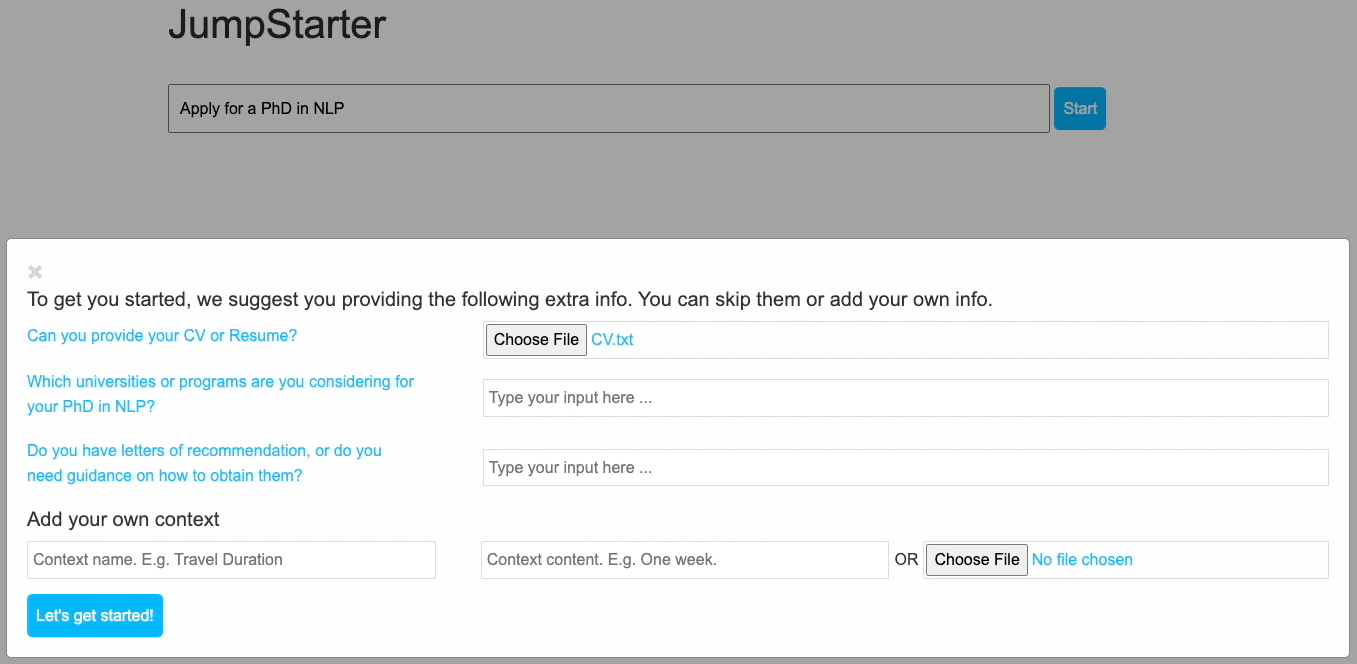}
    \caption{JumpStarter generates questions to elicit context from users to clarify the goal. The user uploads his CV.}
    \label{fig1:gen_ques}
\end{figure}

\begin{figure}[ht]
\centering
    \includegraphics[width=\linewidth]{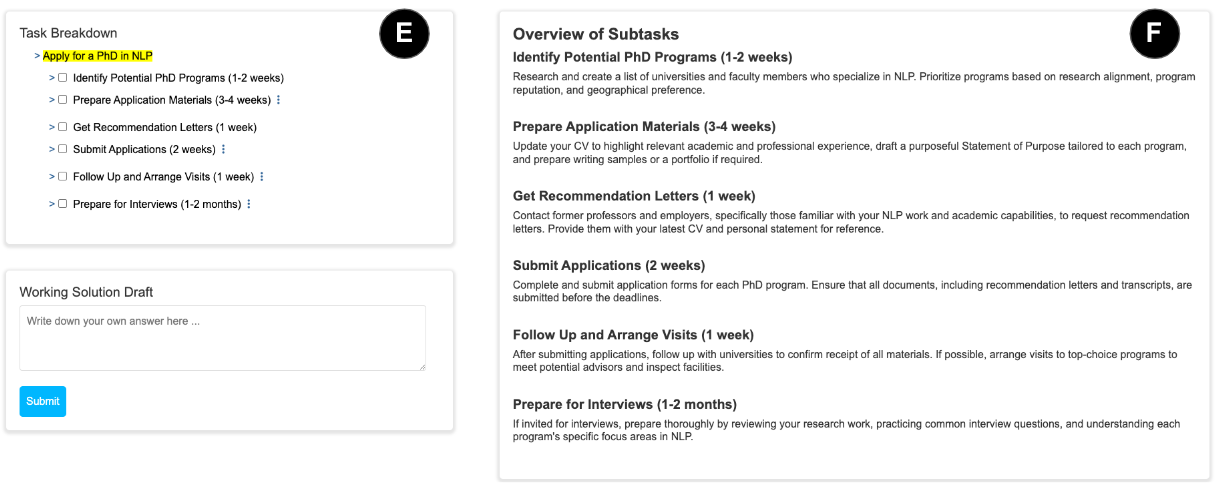}
    \caption{The initial subtask tree and the overview for the goal \textit{Apply for a PhD in NLP}. (E) The task breakdown for the goal. (F) The overview of all the subtasks of the goal, including the titles, descriptions, and duration of completion of the subtasks.}
    \label{fig1:init_tree}
\end{figure}

\begin{figure}[ht]
\centering
    \includegraphics[width=\linewidth]{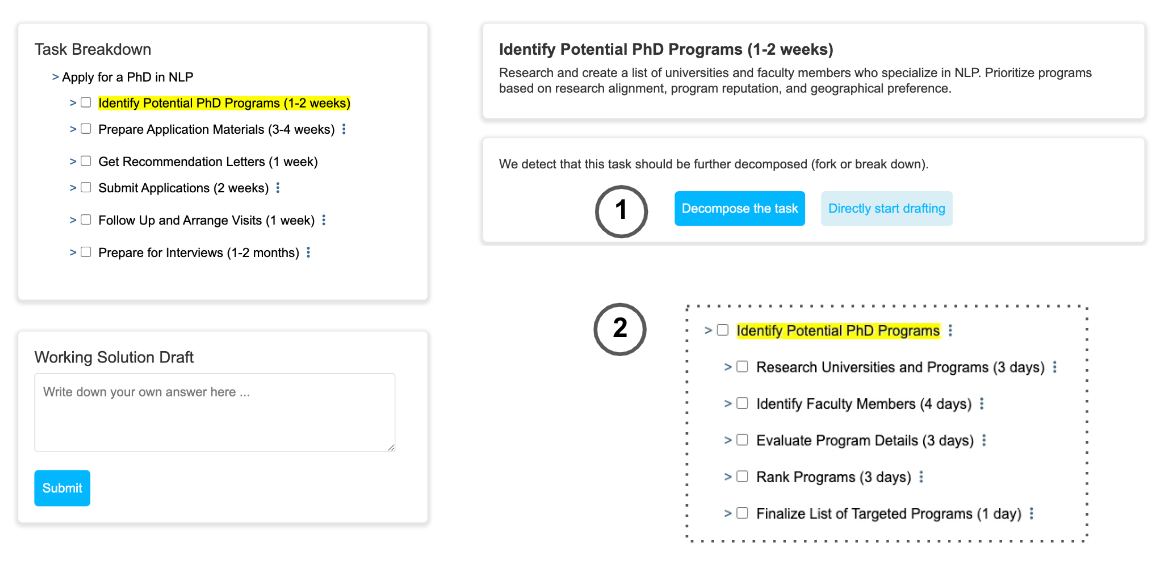}
    \caption{JumpStarter suggests further decomposition for the first subtask \textit{Identify Potential PhD Programs}. (1) John presses the button \textit{Decompose the task}. (2) JumpStarter generates the subtasks for John.}
    \label{fig2:sd_identify}
\end{figure}

\begin{figure}[ht]
    \centering
    \begin{subfigure}[t]{\linewidth}
        \centering
        \vtop{\null \hbox{\includegraphics[width=\linewidth]{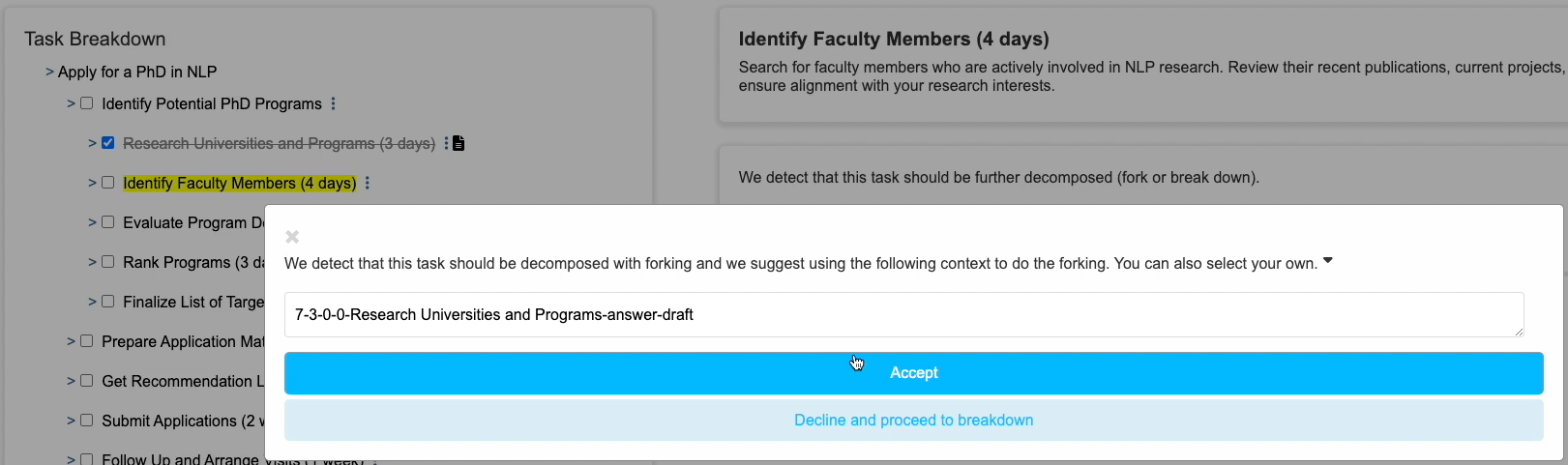}}}
        \caption{Selecting relevant context for forking.}
        \label{fig2:cs_forking}
    \end{subfigure}
    \hfill
    \begin{subfigure}[t]{\linewidth}
        \centering
        \vtop{\null \hbox{\includegraphics[width=\linewidth]{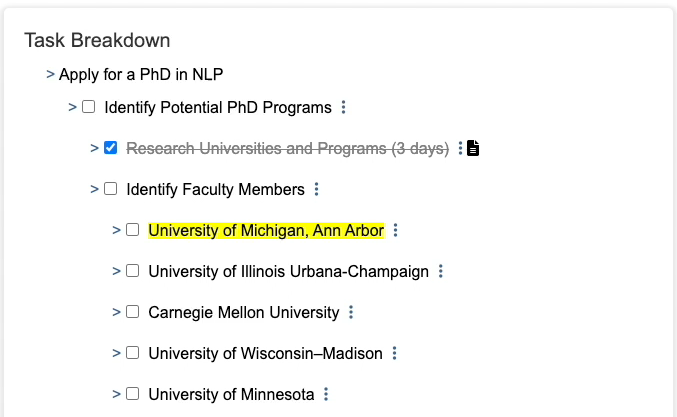}}}
        \caption{Entity-based task decomposition (forking).}
        \label{fig2:nodes_forking}
    \end{subfigure}
    \caption{Context selection for forking and the task decomposition after applying forking on \textit{Identify Faculty Members}.}
    \label{fig:main}
\end{figure}

\begin{figure}[ht]
    \centering
    \begin{subfigure}[t]{\linewidth}
        \centering
        \vtop{\null \hbox{\includegraphics[width=\linewidth]{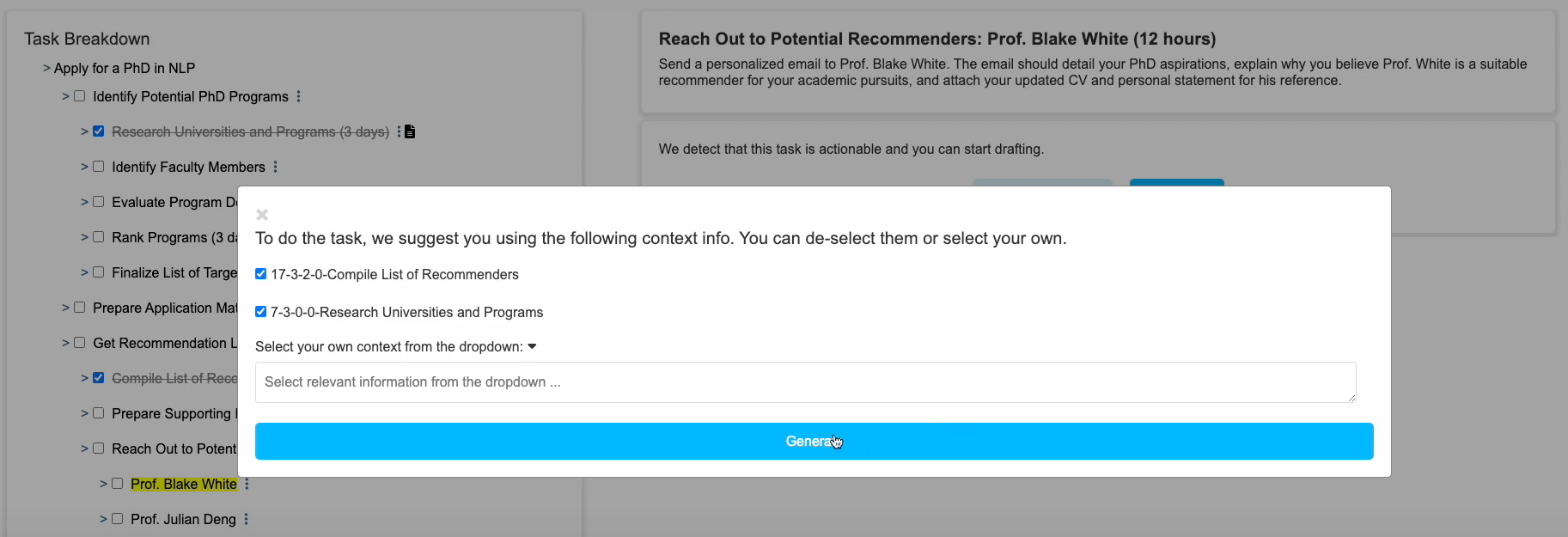}}}
        \caption{Context selection for answer draft generation.}
        \label{fig2:context-selection-drafting}
    \end{subfigure}
    \hfill
    \begin{subfigure}[t]{\linewidth}
        \centering
        \vtop{\null \hbox{\includegraphics[width=\linewidth]{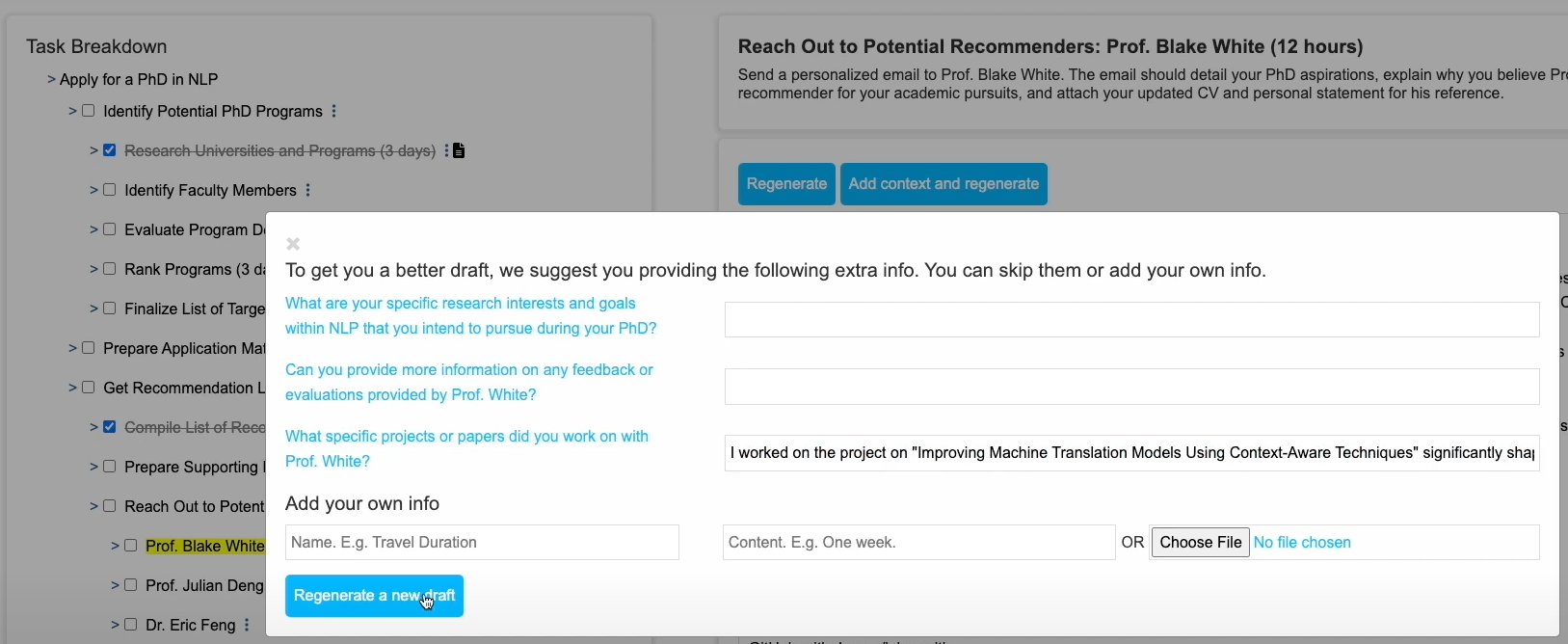}}}
        \caption{Context elicitation for answer draft iteration.}
        \label{fig2:context-elicitation-drafting}
    \end{subfigure}
    \caption{Context selection and elicitation for creating and iterating the answer draft of the subtask \textit{Reach Out to Potential Recommenders: Prof. Blake White.}}
    \label{fig:drafting-main}
\end{figure}

This section presents the interface visuals of the \textit{JumpStarter} system. Figure~\ref{fig:screen_all} shows an overview of the interface as it generates plans and answer drafts for the goal \textit{“Apply for a PhD in NLP”}. Figure~\ref{fig1:gen_ques} displays the elicitation interface used to gather global context before initiating the planning process. Figure~\ref{fig1:init_tree} illustrates the subtask tree, providing a structured overview of the user’s goal. Figure~\ref{fig2:sd_identify} shows how the system recommends further decomposition for a selected subtask. Figure~\ref{fig:main} depicts context selection for task forking and the resulting forked subtask structure. Finally, Figure~\ref{fig:drafting-main} presents the interface for generating an answer draft, including both context selection and additional elicitation.

\section{Details for Pilot Testing}
\label{sec:app_tech_eval}

\subsection{Participants and Procedure}

We validated the design principles of \textit{JumpStarter} through a pilot study. Specifically, we evaluated subtask and answer draft quality for three preselected personal goals inspired by \citet{little1983personal}: (a) Apply to a fellowship, (b) Get a driver’s license, and (c) Organize a team event. 
We then used a university mailing list to recruit participants.
Each goal was assigned two expert participants—those who reported achieving the goal in the past six months. 
Overall, we recruited six expert participants for this study (average age=25.8, three female, three male).
Participants were compensated \$20 per hour, with sessions lasting about 1.5 hours each.

During the study, we first introduced the expert participants to the concept of plans and answer drafts with examples. 
We instructed the participants that they would be using three versions of the system for their respective goals, from start to finish. 
The system versions represented the three experimental conditions, which were presented in shuffled order among participants to counterbalance the learning effect (see Table \ref{tab:expert_participants}). 
We demonstrated how to use each version before each participant used it. 
We asked the participants to generate the subtasks and answer drafts exactly once.
They then rated the perceived quality of the subtasks and answer drafts on a seven-point Likert scale, providing a verbal explanation.

\subsection{Pilot Testing Analysis}

Our results show that the context selection feature significantly enhances the quality of answer drafts compared to the baseline. Unlike context saving only, which keeps all context in the context window all the time, context selection requires the LLM to explicitly choose relevant context from the available pool. 
During sessions under this condition, participants often remarked that the generated solutions appeared to take into account what they had input in previous subtasks. 
This was particularly evident in ``summarizing'' tasks,  where the system could provide a personalized checklist for tasks like applying for scholarships and driver's licenses, or an overall itinerary for a team event. 
E5 referred to the event itinerary they got as a \textit{``very useful synthesis of everything I've explored.''}
In contrast, the context saving only condition tended to produce only general tips for creating the itinerary.
As another example, E1 mentioned that the email draft generated with context selection was more personalized than that created without it. 
E1 stated, \textit{``I like the recommendation letter request email draft it gives me, as it considers much of my background that I saved in the previous `update your CV' task. I did the same thing in the previous round [\textit{context dumping}] but did not feel it was as effective.''} 

\begin{table}[t]
\centering
\begin{adjustbox}{max width=\linewidth}
\begin{tabular}{lll}
\toprule
Participant & Personal goal &Condition order\\ 
\midrule
E1&  \multirow{2}{*}{Apply to a fellowship} & (1) $\rightarrow$ (2) $\rightarrow$ (3) \\ 
E2& & (2) $\rightarrow$ (3) $\rightarrow$ (1)  \\ 
E3&  \multirow{2}{*}{Get a driver's license} &  (3) $\rightarrow$ (1) $\rightarrow$ (2)  \\ 
E4& & (1) $\rightarrow$ (2) $\rightarrow$ (3)   \\ 
E5& \multirow{2}{*}{Organize a team event} & (2) $\rightarrow$ (3) $\rightarrow$ (1)  \\ 
E6& & (3) $\rightarrow$ (1) $\rightarrow$ (2)  \\ 
\bottomrule
\end{tabular}
\end{adjustbox}
\caption{Overview of expert participants for the comparative study. Six experts were assigned one of three goals to evaluate under all three conditions, which were presented in shuffled order to avoid biasing the results.}
\label{tab:expert_participants}
\end{table}

In addition to context selection, context elicitation improves the quality of both subtasks and answer drafts, outperforming both \textit{context dumping} and \textit{context filtering} conditions.
Participants reported that the elicitation questions posed at the beginning \textit{``provided the right plan to start with.''} (E4) 
For instance, E2 uploaded the fellowship requirements document as initially suggested by the system, later rating the generated subtasks a perfect 7/7. 
\textit{``It captured the requirements quite accurately..."} E2 noted. 
\textit{``The subtasks were precise, fitting the unique aspects of the fellowship I am applying to, which requires only one recommendation letter, though typically more are needed.''}  
Similarly, elicitation questions about which state to obtain a driver's license in (for the ``get a driver license'' goal) and how many people are in the team (for the ``organize a team event'' goal) both resulted in subtasks that were better tailored to participants' individual situations.
In addition, preference elicitation questions such as ``What type of vehicle do you intend to drive?'' and ``When do you prefer to hold the event? Weekday or weekend? Noon or night?'' prompted participants to provide answers as personal context, ultimately resulting in more tailored answer drafts that they rated highly.

\section{Evaluation of Subtask Detection}
\label{sec:sub_detect}

\subsection{Experiment settings}
Subtask Detection aims to determine whether a task should be further decomposed to become executable. Using zero-shot and few-shot prompting as baselines, we evaluated three GPT-4-based prompting strategies for identifying actionable subtasks (see Appendix~\ref{sec:prompts} for prompt details). First, we applied Chain-of-Thought (CoT) prompting with few-shot examples~\cite{wei2022chain}. Second, we incorporated the tree level of each task node into the prompt, based on the intuition that higher-level nodes are less likely to be immediately actionable (“Tree”). Third, we explored including the initial answer drafts in the prompt, allowing the model to assess whether decomposition is necessary based on draft quality (“Draft”). We assessed the accuracy of each method using an expert-labeled test suite. All experiments used a temperature of 1, a maximum token limit of 2048, and top-$p$ set to 1. Results are reported as averages across five runs for each setting.

\subsection{Test suite construction}
Drawing inspiration from \citet{little1983personal}, we created a test suite comprising four real-world task
scenarios, each representing a distinct aspect of everyday life: 1) Applying for a PhD program (Academic), 2) Obtaining
a driver’s license (Practical), 3) Finding a surfing camp (Recreational), and 4) Arranging a trip abroad (Travel). For each
scenario, we created ten distinct test cases, resulting in a total of $40$ test cases. An example test case is shown as follows: \textit{Compile a List of Potential Universities: start by identifying the universities that offer PhD programs in Natural Language Processing (NLP). Research and compile a comprehensive list of these universities.}
We recruited four expert participants—one for each task scenario—via a university mailing list (average age=26.3, two
female, two male). These participants reported having completed the tasks in the past six months and were compensated
\$10 for their participation. For each test case, the experts were asked to determine if the current task should be further
decomposed to make it actionable. For the example test case shown above, its label is \textit{"No"} as it is actionable and does not
require further task decomposition.

\subsection{Results and Findings}

\begin{table}
\centering
\begin{adjustbox}{max width=\linewidth}
\begin{tabular}{lrrrr}
\toprule
\multirow{2}{*}{Prompting  Techniques} & \multicolumn{2}{c}{Accuracy} & \multicolumn{2}{c}{Statistics} \\
                                       & Mean          & SD           & $p$              & Sig.           \\ \midrule
Zero-shot                              & .35           & .000         &               &              \\ \midrule
Few-shot                               & .58           & .040         &               &              \\ \midrule
+ CoT                                  & .62           & .050         & .405           &    -           \\
+ CoT + Tree                           & .69           & .020         & .004           &    **           \\
+ CoT + Draft                          & .72           & .020         & .009           &    **          \\
+ CoT + Tree + Draft                                  & .87           & .040         & .000           &      ***         \\ \bottomrule
\end{tabular}
\end{adjustbox}
\caption{The technical evaluation results for Subtask Detection comparing different prompting techniques, where the p-values ($-$: $p>.100$, $+$: $.050<p<.100$, $*$: $p<.050$, $**$: $p<.010$, $***$: $p<.001$) are reported. Note that the p-values are computed against the few-shot-only baseline. Few-shot combined with CoT+Tree+Draft achieved the best accuracy.}
\label{tab:techeval_sd}
\end{table}

As summarized in Table \ref{tab:techeval_sd}, the zero-shot method yielded the lowest accuracy of only 0.35, which also implies an inherent difficulty in the task itself. We observe enhanced performance with few-shot prompting (0.58), with accuracy rising even more when using the CoT prompting paradigm (0.62). Combining CoT few-shot prompting with the task node tree levels (0.69), initial solution draft (0.72), or both (0.87), all significantly led to enhanced performance. Generating the initial solution draft introduced a trade-off between latency and accuracy, so we opted for the slightly less performant few-shot CoT with tree levels for a better user experience.

\begin{table*}[ht]
\centering
\begin{adjustbox}{max width=\linewidth}
\begin{tabular}{llrrrrrrll}
\toprule
\multicolumn{1}{c}{\multirow{2}{*}{\textbf{Category}}}   &
\multicolumn{1}{c}{\multirow{2}{*}{\textbf{Factor}}}               & \multicolumn{2}{c}{JumpStarter}     & \multicolumn{2}{c}{ChatGPT}      & \multicolumn{2}{c}{Statistics} & \multicolumn{1}{c}{\multirow{2}{*}{\textbf{Hypotheses}}}\\
\cmidrule(lr){3-4}\cmidrule(lr){5-6}\cmidrule(lr){7-8}
           &                      & Mean    & SD    & Mean     & SD    &  $p$               & Sig. \\
\toprule

\multicolumn{1}{l}{\multirow{5}{*}{\textbf{Task load}}} & Mental demand & 3.3  & 1.64  & 5.3 & 1.06 
& .005 & ** & \textit{H1a} accepted \\
& Temporal demand & 2.4 & .97 & 4.7     & 1.83  
& .012 & *  & \textit{H1b} accepted     \\
& Performance  & 4.3  & .95  & 2.4  & 1.26  
& .007  & **  & \textit{H1c} accepted    \\
& Effort & 2.8  & 1.69  & 4.8 & 1.23   
& .007 & **  & \textit{H1d} accepted    \\
& Frustration & 2.1 & 1.60 & 4.5 & 1.90   
& .009  & ** & \textit{H1e} accepted   \\
\midrule

\multicolumn{1}{l}{\multirow{2}{*}{\textbf{Exploration efficiency}}}    & Plan items explored  & 4.5 & 1.43 & 2.9 & .74  
& .002  & **  & \textit{H2a} accepted\\
& Answer drafts generated & 5.3  & 1.89 & 3.3 & 1.16 
& .047  & *  & \textit{H2b} accepted \\
\midrule

\multicolumn{1}{l}{\multirow{2}{*}{\textbf{Satisfaction}}}    
& Plan quality & 5.9  & .88 & 3.3 & 1.49 
& .004  & **  & \textit{H3a} accepted \\
& Answer draft quality & 5.8 & .92 & 4.7 & 1.25
& .017 & *  & \textit{H3b} accepted \\
\midrule

\textbf{Confidence} &  & 5.6 & 1.07  & 3.9 & .99  
& .007 & ** & \textit{H4} accepted \\
\bottomrule
\end{tabular}
\end{adjustbox}
\caption{The statistical test results comparing JumpStarter with ChatGPT, where the p-values ($-$: $p>.100$, $+$: $.050<p<.100$, $*$: $p<.050$, $**$: $p<.010$, $***$: $p<.001$) are reported. }
\label{tab:study_result}
\end{table*}

\begin{table*}[t]
\centering
\small
\begin{tabular}{lrrrr}
\toprule
\textbf{Methods} & \textbf{Study Quality} & \textbf{$\Delta$ vs. JS} & \textbf{95\% CI} & \textbf{W/T/L} \\
\midrule
\multicolumn{5}{l}{\textit{\textbf{ChatGPT baselines}}} \\
ChatGPT vanilla & 3.7507 & +0.5192 & [0.4602, 0.5782] & 255/8/37 \\
ChatGPT + elicited context & 3.6992 & +0.5707 & [0.5092, 0.6322] & 257/6/37 \\
ChatGPT + structured summary & 3.7215 & +0.5483 & [0.4897, 0.6070] & 260/6/34 \\
Single-turn decomposition & 4.0503 & +0.2195 & [0.1637, 0.2753] & 197/11/92 \\
\midrule
\multicolumn{5}{l}{\textit{\textbf{Planning and memory-agent baselines}}} \\
ADaPT-style recursive decomposition & 4.0538 & +0.2160 & [0.1649, 0.2671] & 200/20/80 \\
Ask-before-plan & 3.2208 & +1.0490 & [0.9791, 1.1189] & 287/4/9 \\
Memory-RAG & 3.2898 & +0.9800 & [0.9234, 1.0366] & 291/2/7 \\
Integrated agentic planner & 3.9375 & +0.3323 & [0.2813, 0.3833] & 227/14/59 \\
\midrule
\multicolumn{5}{l}{\textit{\textbf{Workflow ablations and context-control baselines}}} \\
Random context selection & 4.1418 & +0.1280 & [0.0690, 0.1870] & 171/17/112 \\
No context selection & 4.0817 & +0.1882 & [0.1275, 0.2488] & 178/9/113 \\
No context reuse & 4.0880 & +0.1818 & [0.1251, 0.2385] & 180/17/103 \\
No elicitation & 4.1903 & +0.0795 & [0.0270, 0.1320] & 163/15/122 \\
\midrule
\textit{JumpStarter-Recursive} & 4.1967 & +0.0732 & [0.0167, 0.1296] & 158/12/130 \\
{\textit{JumpStarter-Shallow}} & \textbf{4.2698} & -- & -- & -- \\
\bottomrule
\end{tabular}
\caption{Automatic simulation results on 300 held-out persona-goal profiles. Study Quality is the blinded judge's composite score. $\Delta$ reports the matched difference between \textit{JumpStarter-Shallow} and each comparator; positive values indicate that JumpStarter performs better. W/T/L reports the number of profiles where \textit{JumpStarter-Shallow} wins, ties, or loses against the methods. \textit{No context selection} passes all available context to every subtask.}
\label{app-tab:auto-results}
\end{table*}

\begin{table}[t]
\centering
\small
\setlength{\tabcolsep}{4pt}
\begin{tabular}{lrrr}
\toprule
\textbf{Profiles} & \textbf{$n$} & \textbf{Loss rate} & \textbf{Gain} \\
\midrule
\multicolumn{4}{l}{\textit{\textbf{Goal complexity}}} \\
Complex & 150 & 31\% & +6.4\% \\
Simple & 150 & 31\% & +4.5\% \\
\midrule
\multicolumn{4}{l}{\textit{\textbf{Time horizon}}} \\
Weeks & 90 & 36\% & +4.7\% \\
Months or longer & 210 & 29\% & +5.8\% \\
\midrule
\multicolumn{4}{l}{\textit{\textbf{Domain}}} \\
Career, education & 80 & 38\% & +4.2\% \\
Creative, personal & 70 & 20\% & +7.4\% \\
Events & 50 & 28\% & +6.1\% \\
Everyday admin & 70 & 33\% & +4.0\% \\
Health, wellness & 30 & 37\% & +6.3\% \\
\midrule
\multicolumn{4}{l}{\textit{\textbf{Persona context richness}}} \\
Low & 45 & 40\% & +2.3\% \\
Medium & 195 & 30\% & +5.7\% \\
High & 60 & 27\% & +7.0\% \\
\bottomrule
\end{tabular}
\caption{Failure-case breakdown of \textit{JumpStarter-Shallow} against single-turn decomposition on the 300 held-out profiles. Loss rate is the share of profiles on which \textit{JumpStarter-Shallow} scores lower; gain is its relative difference in mean study quality. Complex goals are those annotated \textit{high} or \textit{multi-track}; simple goals are \textit{medium}.}
\label{tab:failure-breakdown}
\end{table}

\section{Detailed Design and Results of User Study}
\label{sec:details_user_study}

\subsection{Hypotheses}
In the user study, we investigate the following hypotheses:
\begin{itemize}
    \item \textit{H1}: Compared to ChatGPT, JumpStarter significantly \textbf{lowers users' task load (H1)} for mental demand (H1a), temporal demand (H1b), performance (H1c), effort (H1d) and frustration (H1e).
    \item \textit{H2}: Compared to ChatGPT, JumpStarter significantly \textbf{increases users' exploration efficiency (H2)} in terms of the number of plan items explored (H2a) and the number of answer drafts created (H2b) within the given time.
    \item \textit{H3}: Compared to ChatGPT, JumpStarter significantly \textbf{increases users' satisfaction level with the quality of results (H3)} in terms of plan quality (H3a) and answer draft quality (H3b).
    \item \textit{H4}: Compared to ChatGPT,  JumpStarter significantly \textbf{increases users' confidence in taking the next steps on their personal projects (H4)}.
\end{itemize}


\subsection{Detailed Results and Findings}

\subsubsection{Confidence level in taking the next steps on the goal}
Participants reported significantly higher confidence in taking the next steps on their personal projects using JumpStarter (mean=5.6, SD=1.07) compared to using ChatGPT (mean=3.9, SD=.99).

Participants reported that ChatGPT helped validate their thoughts with commonsense knowledge and sometimes provided surprising or useful tips.
As P7 said, \textit{``Sitting down and planning things out itself is very helpful. I used ChatGPT as a cross-reference, checking to make sure I'm on the right track---thinking similarly to other people. And sometimes answers to low-level tasks covered things I did not really know, which is good.''}
However, they reported that they still always felt they might miss something important while using ChatGPT, whereas with JumpStarter, they feel more secure (P3, P4, P9), as P4 commented, \textit{``I love that I can break things down further if I want, so I don't feel like I miss anything.''}

In addition, JumpStarter can provide very personalized and actionable next steps that greatly increase users' confidence in taking action on their projects.
For example, P1, with the goal of starting a side job, shared that \textit{``The schedule JumpStarter helped me generate is very personalized, and I can directly use it to take real action---before, I felt worried about launching this idea as I had very limited time, now I feel like I can really start doing it.''}
P10 also liked that JumpStarter provided them with a specific and personalized itinerary for organizing the family reunion---\textit{``I like that it summarizes everything I saved in the previous tasks---I can use it in the real world.''}

\subsubsection{Tool preference}
8 out of 10 participants reported they prefer to use JumpStarter in the future, compared to ChatGPT.
The main reasons given include that JumpStarter can provide more customized responses with less cognitive load.
The users feel that they do not have to think hard about what information to provide (P1), are guided by the system (P2), can more easily consume the information (P3) or track the plan (P6), and can get their personal details efficiently organized, framed, and utilized (P8, P9, P10).

The other two participants (P4 and P5) mentioned that their choice depended on how familiar they were with the project they wanted to work on.
If it was a topic they already had a clear understanding of, they preferred the chatbot interaction to help them figure out the details. 
Otherwise, they would prefer to use JumpStarter as it offers more structure. 

\subsubsection{Improvement feedback}
Participants also provided insights on how to improve JumpStarter. 
The main feedback included ``\textit{make the subtask outline and task descriptions editable}'' (P1, P3, P6, P7, P8, P9), ``\textit{format the suggested answer draft to be easier to read}'' (P4, P8, P9), ``\textit{enable users to add or edit context whenever they want}'' (P3, P7), and ``\textit{add a synthesis button to summarize what has been explored so far}'' (P1, P2). 
P8 also suggested embedding a search agent to collect data and ensure credibility.
We discuss limitations and future work further in Section ~\ref{sec:limitations}.

\subsection{Additional User Quotes from User Study}
\subsubsection{\textit{JumpStarter} Reduces Users' Task Load}

As shown in Figure \ref{fig:nasa}, in the NASA TLX dimensions~\cite{nasa}, working with \textit{JumpStarter} was significantly less demanding in mental demand (p=.005), temporal demand (p=.012), performance (p=.007), effort (p=.007), and frustration (p=.009). 

Participants attributed this to the structured interface and context-aware guidance of \textit{JumpStarter}, which helped reduce cognitive load. 
Unlike ChatGPT’s linear chat format, \textit{JumpStarter} made it easier to visualize progress and maintain task structure. 
As P2 noted, \textit{``ChatGPT info dumps a lot, and I have to keep the structure in my brain, whereas JumpStarter gave me a structure that I could easily follow.''}
Similarly, targeted questions helped users refine their input and move forward. 
P10 commented, \textit{``I appreciate the questions JumpStarter asked when I felt stuck about how to iterate the answer draft.''}
In contrast, ChatGPT required users to generate and manage context manually. P8 remarked, \textit{``In ChatGPT, the information load is high—I have to think very hard about what info I should provide to get things that work for me.''}

Participants also reported expending more effort with ChatGPT, often without meaningful improvement in output quality.
For example, P1 asked ChatGPT to generate clarifying questions to improve the answer draft. 
It returned eight, which P1 described as \textit{``a bit too abstract and hard to answer.''} 
Despite answering them all, P1 felt the revised output remained \textit{``too general and not useful.''}
They also noted that \textit{``ChatGPT seemed to forget these eight answers soon after''}, leading to frustration and a sense that their effort was wasted.

\subsubsection{\textit{JumpStarter} Enables Better Exploration Efficiency}
Participants explored significantly more plan items using \textit{JumpStarter} (mean=4.5, SD=1.43) than with ChatGPT (mean=2.9, SD=0.74), and created more answer drafts as well (mean=5.3, SD=1.89 vs. mean=3.3, SD=1.16). Figure~\ref{fig:exploration} illustrates this difference.

One reason for this disparity is that ChatGPT’s linear interaction format often leads users to fixate on a single task, limiting broader exploration. 
For instance, P2 iterated nine times on one answer draft: \textit{``It took nine iterations to get the draft I like. I really hoped ChatGPT would guide me, but I had to direct myself. I got so involved that I completely forgot I had other planning items.''}

In contrast, \textit{JumpStarter}’s task-structured interface helped users stay oriented within the broader plan while focusing on one task at a time. 
P4 remarked, \textit{``JumpStarter has a more flexible structure. I like that I can easily jump between tasks... Seeing the task description and relevant context gives me everything I need''} 
P10 similarly noted, \textit{``JumpStarter automatically manages and considers my drafts from previous tasks, which is great and helps me focus on the current one.''}

\subsubsection{\textit{JumpStarter} Improves Perceived Quality of Outcome}
Plans created with \textit{JumpStarter} were rated significantly higher in quality (mean=5.9, SD=0.88) than those created with ChatGPT (mean=3.3, SD=1.49).
Similarly, participants rated the quality of answer drafts higher with \textit{JumpStarter} (mean=5.8, SD=0.92) compared to ChatGPT (mean=4.7, SD=1.25). 
Figure~\ref{fig:satisfaction} shows these differences in perceived quality.

Participants attributed this improvement to \textit{JumpStarter}’s proactive questioning and personalized planning.
Unlike ChatGPT, which often produced generic plans, \textit{JumpStarter} asked targeted questions early in the process and incorporated user responses into the plan structure. 
For example, P9, who was planning to start a YouTube channel, noted: \textit{``JumpStarter asked if I was experienced with video editing, and I said yes. It was reflected in the plan accordingly, unlike ChatGPT, which focused too much on editing I didn’t need.''}

In addition, participants also appreciated how \textit{JumpStarter} effectively uses the relevant context to personalize the answer draft, as P1 noted, \textit{``JumpStarter gave me much more tailored responses—like a personalized schedule to help me start my side job. It took into account key details like my limited time and the specific area I am interested in. It handled the context very well.''}

\subsection{Extended Discussion about User Study}
\label{sec:ext_dis}

\textit{JumpStarter} demonstrates how adaptive personal context curation through context elicitation, condensation, and reuse can enhance human-AI collaborative planning by improving both plan quality and user experience. Unlike ChatGPT’s user-driven \textit{“pull”} model, \textit{JumpStarter}’s \textit{“push”} approach reduces cognitive load by guiding users with targeted questions and enabling selective context previews, aligning with recognition-over-recall principles from cognitive psychology~\cite{craik1972levels}. Users valued this structured interaction but also expressed a desire for more persistent and adaptive memory, particularly as their goals and preferences evolved, highlighting the need for systems that can update context dynamically over time~\cite{wang2024ai}. 

While ChatGPT is designed to retain conversational context, it often failed to apply it meaningfully, occasionally generating contradictory responses. These findings underscore the importance of designing systems that support more seamless and transparent context management. Looking forward, integrating in-situ context capture from tools like email or desktop files~\cite{umea, programmer_context, bergman2000making} may enable more fluid, real-world human-AI collaboration. \textit{JumpStarter} offers a design paradigm for building systems that better align LLM capabilities with users' evolving needs in complex, long-term planning workflows.

\subsection{Scope of the User Study}
With ten participants in one session each, the study establishes the direction of the effect for real users rather than its size or generality. All participants were college students already familiar with ChatGPT; this familiarity favors the baseline, since \textit{JumpStarter} was new to every participant. The design principles were validated separately in a pilot study with six expert evaluators (\S\ref{sec:pilot-design}), and the simulation study (\S\ref{sec:simulation}) tests whether the effect holds across a broader population of simulated users, including novice, low-context, terse, and deferential personas.

\section{Details on Dataset Construction for Automatic Simulation}
\label{app:details_dataset_con}

The dataset for the automatic simulation experiment is constructed in three deterministic
stages: (i)~a 60-goal benchmark grounded in 13 anchor sessions from the human study, (ii)~a 20-persona set extracted from the same study traces and augmented with controlled behavioral variants, and (iii)~a $1,200$-profile persona\,$\times$\,goal cross-product that fixes the private state each simulated user holds during a session. 

\subsection{Realistic Goal Synthesis}
\label{sec:appendix-goal-synthesis}

\subsubsection{Anchor Sessions}
\label{sec:appendix-anchor-sessions}

We first extract 13 anchor sessions from the JumpStarter human study: 10~\emph{human-study traces} from participants P1--P10 (one per participant, each participant's self-chosen planning goal) and 3 traces drawn from the pilot study tasks. Each anchor session records the participant identifier, goal text,
domain, the participant's saved global and user context, background snippets, saved local-context entries, and saved drafts. 

\subsubsection{Domain Distributions}
\label{sec:appendix-domain-quotas}

Goals are stratified across five planning domains with fixed quotas that sum to 60 (Table~\ref{tab:appendix-domain-quotas}). 
The goal set covers five domains: career and education, creative and personal projects, everyday administration and home planning, event coordination, and health/wellness/training. Expected outputs include schedules, checklists, comparison matrices, outreach drafts, study plans, and budgets. The synthesized goal set is designed to avoid tasks where a generic one-shot response would be sufficient.


\begin{table}[t]
\centering
\small
\begin{tabular}{lr}
\toprule
Domain & Count \\
\midrule
\texttt{career\_education}        & 16  \\
\texttt{creative\_personal}       & 14 \\
\texttt{everyday\_admin\_home}    & 14  \\
\texttt{events\_coordination}     & 10  \\
\texttt{health\_wellness\_training}&  6  \\
\midrule
\textbf{Total}                    & \textbf{60} \\
\bottomrule
\end{tabular}
\caption{Domain distributions for the 60 goals.}
\label{tab:appendix-domain-quotas}
\end{table}

\subsubsection{Goal Synthesis Prompt and Model}
\label{sec:appendix-synthesis-prompt}

Goals are synthesized with \textbf{GPT-5.5}. The prompt is shown below:
\hfill
\begin{tcolorbox}[promptbox, title={Prompt for Goal Synthesis},fonttitle=\small]
You generate human-grounded planning-goal benchmark candidates for evaluating task-structured context curation.

Return only structured data matching the provided schema. Each goal must be realistic for an individual user, require personal context, benefit from decomposition, and support tangible outputs such as schedules, drafts, checklists, itineraries, application materials, or decision aids.

Write `goal\_text` as a short, plain goal title, not a full scenario sentence. It must be easy to parse at a glance: \\
- 3-9 words.
- 70 characters or fewer.
- No final period.
- No commas, semicolons, colons, or multi-clause wording.
- Avoid clauses such as "while...", "within...", "with...", "for exploring...", "that...".
- Put details and constraints in `known\_context\_needs`, `expected\_tangible\_outputs`, `time\_horizon`, and other metadata fields instead. \\

Good goal\_text examples:\\
- Prepare for the GRE exam
- Apply for an NSF Graduate Research Fellowship
- Organize a PhD welcome picnic
- Prepare for the bar exam
- Start a side job
- Design a personal website
- Start a personal YouTube channel
- Design a wardrobe for four seasons
- Move to a new apartment
- Prepare for the half-marathon
- Organize a family reunion party at home \\

Avoid goals that are mostly fact lookup, one-shot generic advice, unsafe, too verbose, hard to parse, or too close to the anchor goals. Match the requested domain counts exactly.

\end{tcolorbox}



\subsubsection{Filtering and Quality Verification on Synthesized Goals}
\label{sec:appendix-filtering}

Each candidate passes through two filters before acceptance. A
{structural gate} requires both
\texttt{requires\_personalization} and \texttt{requires\_decomposition}
to be true, non-empty \texttt{expected\_tangible\_outputs} and
\texttt{known\_context\_needs}, compliance with the style rubric, and 
token-overlap similarity below $0.8$ against every accepted goal. A separate GPT-5.5 {LLM filter} re-evaluates each candidate against the same rubric (see prompt below); acceptance
requires both filters to pass. The synthesis--filter loop runs up to
four times until 60 goals are
accepted. Finally, the leading authors manually inspected all 60
goals end-to-end, confirming realism, decomposability, and the
absence of duplicates or anchor-leakage before finalizing the goal set.

\begin{tcolorbox}[promptbox,title={Prompt for Goal Filtering},fonttitle=\small]
You are filtering one planning benchmark candidate.

Accept only if all checks are true:
- realistic for an individual user,
- requires personal context,
- benefits from decomposition,
- can produce tangible artifacts,
- is not mostly a search/fact-lookup task,
- is not too similar to the nearest existing goal.
- has a short, plain `goal\_text` that reads like a goal title rather than a full scenario sentence.

Reject if `goal\_text` is over 9 words, over 70 characters, has clause punctuation, ends with a long constraint clause, or uses phrasing such as "while...", "within...", "with...", "for exploring...", or "that...".

Return only structured data matching the schema.

\end{tcolorbox}

\subsection{Simulated Users Construction}
\label{app:sim_users}

The simulated-user set is the cross-product of a 20-persona set
and the 60 synthesized goals ($1,200$ \emph{simulation profiles}).
Personas describe \emph{who the user is}; profiles bind a persona to
a specific goal and encode the \emph{private state} the user holds
during a session: what they know, what they have not decided, what
they will or will not disclose, what they will approve, and when they
will push back.

\subsubsection{Base Persona Extraction}
\label{sec:appendix-persona-extraction}

We use \textbf{GPT-5.5} to extract the 10 base personas from the
human-study anchors of P1--P10, one persona per participant. The
extraction prompt (shown below) runs
under structured output and is conditioned on the participant's
saved global context, user context, background, and saved drafts;
the model is required to summarize behavior, constraints, and
communication style, and to ground
every persona in evidence quotes from the anchor. Each
persona is characterized along four behavioral axes:
\begin{itemize}\setlength\itemsep{2pt}
  \item \textbf{Expertise level}~$\in$~\{novice, intermediate, expert\}, inferred from the participant's stated background and the sophistication of saved drafts.
  \item \textbf{Context richness}~$\in$~\{low, medium, high\}, inferred from the density of saved global and user-context entries.
  \item \textbf{Decision style}~$\in$~\{opinionated, deferential, collaborative\}, inferred from the presence and forcefulness of stated preferences and constraints.
  \item \textbf{Communication style}~$\in$~\{terse, balanced, verbose\}, inferred from sentence length and bullet density in the saved drafts.
\end{itemize}
Alongside these axes, the model produces \texttt{task\_objectives},
\texttt{known\_facts}, \texttt{preferences},
\texttt{constraints\_to\_remember}, \texttt{gaps\_in\_self\_knowledge},
\texttt{information\_disclosure\_policy},
\texttt{response\_style\_rules}, \texttt{success\_criteria},
\texttt{pushback\_triggers}, and \texttt{evidence\_quotes}.
Identifiers (\texttt{persona\_id}, \texttt{source\_participant},
\texttt{source\_anchor\_id}, \texttt{goal}) and evidence quotes are
preserved verbatim. To prevent identifiable
institutional information from leaking into model-facing fields, all
references to the participants' real universities are masked out.

\begin{tcolorbox}[promptbox,title={Prompt for Persona Extraction},fonttitle=\small]
You refine persona profiles for a human-AI planning benchmark.

Each persona should describe behavior, constraints, and communication style, not final task answers. Preserve grounding in evidence quotes and do not invent biographical details beyond what the evidence supports.

Return only structured data matching the provided schema. Preserve:
- persona\_id
- persona\_type
- source\_participant
- source\_anchor\_id
- goal
- evidence\_quotes \\

You may improve:
- user\_description
- expertise\_level
- context\_richness
- decision\_style
- communication\_style
- task\_objectives
- known\_facts
- preferences
- gaps\_in\_self\_knowledge
- constraints\_to\_remember
- information\_disclosure\_policy
- response\_style\_rules
- success\_criteria
- pushback\_triggers
- extraction\_warnings \\

Do not include ablation anchors as human-study personas.

\end{tcolorbox}


\subsubsection{Controlled Persona Variants}
\label{sec:appendix-persona-variants}

To stress-test the workflow under behaviorally distinct users while
keeping each variant grounded in a real anchor, we generate 10 additional \emph{controlled variants}. Each variant copies a base persona and toggles a single behavioral axis to a contrastive value;
\texttt{variant\_axis}, \texttt{variant\_description}, and an
extraction warning of the form \emph{``Controlled variant modifies
$\langle$axis$\rangle$ from base persona''} are recorded. The full
variant specification is given in Table~\ref{tab:appendix-persona-variants}. The final persona set is
therefore 20 personas (10~base + 10~variants).

\begin{table}[t]
\centering
\small
\begin{tabular}{lll}
\toprule
Source & Axis & Value \\
\midrule
P2  & context\_richness     & low           \\
P3  & context\_richness     & high          \\
P4  & decision\_style       & deferential   \\
P5  & decision\_style       & opinionated   \\
P6  & communication\_style  & terse         \\
P7  & communication\_style  & verbose       \\
P8  & expertise\_level      & novice        \\
P9  & expertise\_level      & intermediate  \\
P10 & context\_richness     & medium        \\
P1  & decision\_style       & collaborative \\
P6  & communication\_style  & balanced      \\
\bottomrule
\end{tabular}
\caption{Controlled-variant specification: each variant clones a base
persona and toggles a single behavioral axis to the listed value.}
\label{tab:appendix-persona-variants}
\end{table}

\subsubsection{Persona Validation and Example}
\label{sec:appendix-persona-validation}

Persona validation enforces: (i)~exactly 10 base personas and
10 controlled variants; (ii)~the set of base
\texttt{source\_participant} identifiers equals
$\{\textsc{P1},\dots,\textsc{P10}\}$; (iii)~every \texttt{persona\_id}
is unique; (iv)~every persona has at least one evidence quote;
(v)~every controlled variant has a non-null \texttt{variant\_axis};
(vi)~\texttt{constraints\_to\_remember} is non-empty (a warning
otherwise). Below is an example persona:

\begin{small}
\begin{verbatim}
{
  "persona_id": "P2_base",
  "persona_type": "base",
  "source_participant": "P2",
  "goal": "Organize a weekly PhD game night",
  "expertise_level": "novice",
  "context_richness": "medium",
  "decision_style": "opinionated",
  "communication_style": "balanced",
  "task_objectives": [
    "coordinate people, time, and location",
    "create a feasible event plan",
    "draft communication or logistics artifacts"
  ],
  "known_facts": [
    "Never organized game night before",
    "Availability: after 7:30pm, weekend nights",
    "Interests: STEM PhDs, fantasy/party games",
    "Size: 5-9 PhD students",
    "University: Cedarwick Hollow University"
  ],
  "constraints_to_remember": [
    "Interests: STEM PhDs, fantasy/party games",
    "Never organized game night before"
  ],
  "information_disclosure_policy": [
    "Do not volunteer unstated details; reveal
     known facts only when asked.",
    "If asked about absent info, say the user
     does not know yet or has not decided.",
    "Keep answers brief; let the assistant
     elicit missing planning context."
  ],
  "response_style_rules": [
    "Answer in 1-3 concise sentences.",
    "Provide specifics when asked; do not
     front-load every detail.",
    "Use a collaborative tone."
  ],
  "evidence_quotes": ["...verbatim spans..."]
}
\end{verbatim}
\end{small}

\subsubsection{Simulation Profiles}
\label{sec:appendix-profiles}

A simulation profile combines one persona with one benchmark goal,
yielding $20 \times 60 = 1{,}200$ profiles in total. Each profile
specifies the private state the simulated user holds at the start of
a session: \texttt{private\_context\_facts} the user knows but does
not necessarily volunteer; \texttt{unknown\_or\_undecided\_facts}
(e.g., budget, deadline, audience); \texttt{must\_reveal\_if\_asked}
and \texttt{do\_not\_volunteer} sets that govern disclosure;
\texttt{approval\_criteria} and \texttt{pushback\_rules} that govern
acceptance; the frozen \texttt{stable\_persona\_traits} block; and
\texttt{turn\_style} parameters for multi-turn behavior. In
\emph{source-goal pairs} the profile inherits the persona's
\texttt{known\_facts} and \texttt{constraints\_to\_remember} as
private context; in \emph{cross-goal pairs} biographical specifics
are dropped and only \texttt{stable\_persona\_traits} carries over,
preventing persona-specific facts (e.g., ``900+ followers'') from
leaking into unrelated goals. Profiles are produced by \texttt{GPT-5.5} (see prompt below), with identifiers
preserved verbatim and no biographical invention permitted.

A validation run checks that all $1{,}200$ persona--goal pairs
are present, that \texttt{simulation\_profile\_id} values are
unique, and that every profile has non-empty
\texttt{approval\_criteria} and \texttt{pushback\_rules} and a
positive \texttt{turn\_style.max\_tokens}. 

\begin{tcolorbox}[promptbox,title={Prompt for Profile Generation and Validation},fonttitle=\small]
You refine private simulation profiles for a persona-grounded user simulator.

Each profile combines one persona with one benchmark goal. Preserve:
- simulation\_profile\_id
- persona\_id
- goal\_id
- goal\_text
- domain
- source\_persona\_summary \\

You may improve:
- private\_context\_facts
- unknown\_or\_undecided\_facts
- must\_reveal\_if\_asked
- do\_not\_volunteer
- approval\_criteria
- pushback\_rules
- stable\_persona\_traits
- turn\_style
- initial\_state
- state\_tracking\_notes \\

Rules:
- Do not invent biographical facts not supported by the source\_persona\_summary or goal metadata.
- It is okay to add goal-specific unknowns such as budget, deadline, availability, target audience, or prior experience when the benchmark goal implies them.
- Keep facts concise and simulator-actionable.
- Make the profile useful for multi-turn state tracking.
- Return only structured data matching the schema.

\end{tcolorbox}


An example simulation profile is shown below. At simulation time, this profile, the persona block, and the goal
record are rendered into the GPT-4o user-simulator prompt. The
simulator's turn-by-turn behavior---whether to volunteer a fact,
whether to push back, whether to accept a draft, whether to expand a
node---is governed by the disclosure policy,
\texttt{must\_reveal\_if\_asked}, \texttt{do\_not\_volunteer},
\texttt{pushback\_rules}, and the calibrated decision policies of
\S\ref{sec:appendix-decision-calibration}.

\begin{small}
\begin{verbatim}
{
  "simulation_profile_id": "P2_base_G001",
  "persona_id": "P2_base",
  "goal_id": "G001",
  "goal_text": "Organize a weekly PhD game night",
  "domain": "events_coordination",
  "private_context_facts": [
    "Never organized game night before",
    "Availability: after 7:30pm, weekend nights",
    "Interests: STEM PhDs, fantasy/party games",
    "Size: 5-9 PhD students",
    "University: Cedarwick Hollow University",
    "Required outputs: venue shortlist,
     semester calendar, invitation message",
    "Time horizon: multi-week"
  ],
  "unknown_or_undecided_facts": [
    "expected group size: unknown until asked",
    "campus location: unknown until asked",
    "preferred game types: unknown until asked"
  ],
  "must_reveal_if_asked": [
    "Interests: STEM PhDs, fantasy/party games",
    "Never organized game night before",
    "expected group size",
    "campus location constraints",
    "preferred game types"
  ],
  "do_not_volunteer": [
    "expected group size",
    "campus location constraints",
    "preferred game types",
    "Size: 5-9 PhD students",
    "Availability: after 7:30pm, weekend nights"
  ],
  "approval_criteria": [
    "plan directly supports the stated goal",
    "assistant uses stated context only",
    "assistant asks for missing information
     before finalizing details",
    "includes venue shortlist",
    "includes semester calendar",
    "includes invitation message"
  ],
  "pushback_rules": [
    "push back on contradicted constraints",
    "push back on invented facts",
    "push back on generic plans w/o elicitation",
    "push back on skipped personal-context asks",
    "push back on missing tangible artifacts",
    "push back on fact lookup or one-shot advice"
  ],
  "stable_persona_traits": {
    "expertise_level": "novice",
    "context_richness": "medium",
    "decision_style": "opinionated",
    "communication_style": "balanced",
    "response_style_rules": [
      "Answer in 1-3 concise sentences.",
      "Provide specifics when asked; do not
       front-load every detail.",
      "Use a collaborative tone."
    ]
  },
  "turn_style": {
    "max_tokens": 220,
    "register": "collaborative",
    "length": "short"
  }
}
\end{verbatim}
\end{small}

\subsubsection{User Simulator Calibration}
\label{sec:appendix-decision-calibration}

We calibrate simulated-user behavior using 16 logged \textit{JumpStarter} sessions from the user study and pilot testing to reduce over-compliance in the automatic benchmark. From these traces, we estimate empirical behavior rates for task-tree acceptance, subtask decomposition, draft saving, node completion, context provision, revision, and final-readiness decisions (Table~\ref{tab:appendix-decision-rates}). The logs show that users generally accepted the top-level task structure, with an average of 5.73 top-level children, but were more selective about deeper decomposition: users proceeded with further decomposition in 72.1\% of detected opportunities under a broad proxy, while realized child decomposition occurred in only 12.2\% of cases. Users completed 53.2\% of nodes on average and saved drafts for 51.0\% of nodes.

We validate the calibrated simulator with leave-one-session-out diagnostics across ten binary behavior families. The simulator achieves a mean Brier score of 0.157, five-bin expected calibration error of 0.047, and session-level absolute rate error of 0.214, indicating that its predicted decision probabilities are reasonably aligned with held-out real-user behavior. Thresholding decisions at 0.5 yields mean accuracy of 0.809 and mean F1 of 0.898. These results suggest that the calibrated simulator better matches empirical interaction patterns from real \textit{JumpStarter} sessions, instead of assuming uniformly cooperative users who always accept decompositions, save drafts, complete nodes, or approve final outputs.

\begin{table}[t]
\centering
\scriptsize
\setlength{\tabcolsep}{2pt}
\renewcommand{\arraystretch}{1.0}
\begin{tabularx}{\columnwidth}{@{}Xc@{}}
\toprule
\textbf{Decision point} & \textbf{Rate ($n$)} \\
\midrule
Accept top-level task tree & $1.000$ (16) \\
Subtask: decompose further / draft & $.721/.279$ (197) \\
Realized child decomposition ($\geq$ depth 2) & $.122$ (197) \\
Save a generated draft & $.748$ (107) \\
Node marked completed & $.776$ (107) \\
Multi-generation revision before save & $.832$ (107) \\
Session provides global context & $.733$ (16) \\
Session provides local context & $.933$ (16) \\
\bottomrule
\end{tabularx}
\caption{Empirical base rates from human-study logs used as simulator decision policies.}
\label{tab:appendix-decision-rates}
\end{table}

\section{LLM-based Context Relevance Labeler}
\label{app:ctxt_label}

To produce evidence on whether baseline methods select and reuse
\emph{relevant} context, we score every available context item
against every subtask with an LLM-based relevance labeler. For each
\texttt{context\_selection} event in a session trace, the labeler emits one judgment per (subtask, available context item) pair: it
sees only the subtask name and description and the context item's
label and value, and produces a relevance score. 

We use \texttt{GPT-5.4-mini} to judge the context relevance. The prompt is shown below. Specifically, relevance is judged \emph{locally}: the model is told to score the relationship between the specific subtask and the specific context item. Besides, the score-to-decision mapping is explicit:
\texttt{[0.00,0.20]} marks irrelevant or distracting items,
\texttt{[0.21,0.40]} weak background,
\texttt{[0.41,0.70]} supporting context that could help, and
\texttt{[0.71,1.00]} directly useful context; the binary
\texttt{relevant} flag must be \texttt{true} only when the score is
at least $0.50$ \emph{and} the value is concrete enough to act on. 

\begin{tcolorbox}[promptbox,title={Prompt for Context Relevance Labeler},fonttitle=\small]
You are labeling whether a saved user-context item is useful for a specific planning subtask.

Evaluate only the relationship between the subtask and the context item. Do not reward a context item merely because it is generally related to the overall goal. A relevant context item should help
the system make a concrete decision, fill a field, personalize an artifact, avoid a mistake, or reuse prior work for this specific subtask. \\

Use this scale: \\
- 0.00-0.20: irrelevant, distracting, or unusable \\
- 0.21-0.40: weak background context \\
- 0.41-0.70: supporting context that could help the subtask \\
- 0.71-1.00: directly useful context for the subtask \\

Mark relevant=true only when relevance\_score is at least 0.50 and the context is concrete enough to use. If the context says the user does not know, is undecided, or gives no actionable detail, mark relevance\_type="unknown\_or\_unusable" and relevant=false even if the label sounds related.

Return strict JSON matching the schema.

\end{tcolorbox}

\section{Details on LLM-as-a-Judge for Automatic Simulation Evaluation}
\label{app:llm-as-judge}

\subsection{Rubric and Prompt}
\label{sec:appendix-judge-rubric}

Every artifact is scored on a 1--7 Likert scale matching the human
study, anchored as \emph{1\,=\,very poor or unusable},
\emph{4\,=\,adequate}, \emph{7\,=\,excellent, highly actionable, and
well grounded}. The judge (i.e. \texttt{GPT-5.5}) fills a strict-JSON
\texttt{JudgeRubricScore} with the following criteria, each carrying an
integer score and a short evidence string:
\begin{itemize}\setlength\itemsep{1pt}
\item \texttt{plan\_quality} --- coherent, useful plan for the goal.
\item \texttt{tangible\_result\_quality} --- directly usable artifacts (drafts, schedules, checklists, trackers, templates, messages, decision aids), not advice.
\item \texttt{confidence\_support} --- helps the user feel ready to take real-world action.
\item \texttt{personalization\_context\_grounding} --- correct use of user context, respecting constraints.
\item \texttt{completeness\_coverage} --- covers the important subtasks.
\item \texttt{decomposition\_quality} --- breaks the goal into subtasks, with sequencing and dependencies.
\item \texttt{context\_curation\_reuse} --- elicits, preserves, selects, or reuses user context across subtasks.
\item \texttt{workflow\_progress\_support} --- supports progress across the planning workflow, including partial completions.
\item \texttt{user\_burden\_reduction} --- reduces steering, re-explaining, and cleanup the user must do.
\item \texttt{specificity} --- concrete and tailored rather than generic.
\item \texttt{no\_contradiction\_hallucination} --- avoids contradictions, unsupported assumptions, and fake specificity.
\item \texttt{overall} --- holistic score.
\end{itemize}

The final score for each method in Table~\ref{tab:auto-results} is the weighted average of the first nine criteria: 0.15 each for plan quality, tangible-result quality, and confidence support; 0.10 each for personalization and context grounding, completeness, decomposition quality, context curation and reuse, and workflow-progress support; and 0.05 for user-burden reduction. Specificity, contradiction avoidance, and the overall score are recorded as diagnostics and are not part of the composite.

The prompt for the LLM-judge is shown below. Specifically, judging is a two-step pipeline. First, an \textit{evidence-extraction} call returns a \texttt{JudgeExtractedEvidence} record. This record
is attached to the artifact and passed to the final scoring prompt as an audit aid. Second, the scoring prompt is conditioned on
the goal, the user context, the raw artifact, and the extracted
evidence, and emits the \texttt{JudgeRubricScore} above.

\begin{tcolorbox}[promptbox,title={Prompt for LLM-Judge Evidence Extraction},fonttitle=\small]
You are extracting evidence to help evaluate an AI-generated planning artifact.

Use the user context and raw artifact exactly as given. Extract evidence symmetrically for any system or condition; do not favor workflow-style outputs or single-response outputs. \\

Rules: \\
- Do not score quality here. \\
- Do not invent missing details. \\
- Include concise concrete evidence from the artifact. \\
- If evidence is absent, use an empty list. \\
- Treat placeholders, TODO fields, unsupported specifics, contradictions, and cleanup burden as important evidence. \\
- A workflow artifact can show value through task decomposition, saved context, context reuse, and partial tangible drafts, but clutter and stale process notes are also evidence. \\
- A clean prose artifact can show value through directly usable final outputs, but it should not receive credit for context reuse or workflow memory unless the artifact actually shows it. \\

Return strict JSON matching the schema with these fields: \\
- tangible\_outputs\_found: concrete drafts, schedules, checklists, trackers, surveys, messages, templates, decision aids, or other usable artifacts. \\
- confirmed\_user\_context\_used: user-provided facts or constraints from user\_context that the artifact uses correctly. \\
- unresolved\_todos\_or\_placeholders: TODOs, bracketed placeholders, missing owners/dates/details, or unresolved open decisions. \\
- unsupported\_or\_invented\_details: specific names, capacities, deadlines, venues, claims, policies, or personal details that appear unsupported by user\_context or are contradicted. \\
- workflow\_progress\_evidence: decomposition, sequencing, partial completions, saved drafts, or multi-step progress shown in the artifact. \\
- context\_reuse\_evidence: evidence that context was elicited, preserved, selected, or reused across subtasks. \\
- user\_burden\_evidence: evidence that the user would need more steering, cleanup, verification, re-explaining, or manual reconciliation. \\
- final\_actionability\_summary: one concise sentence summarizing how ready the artifact is for real-world action. \\
- extraction\_warnings: caveats about ambiguity, truncation, or evidence that could not be determined. \\

\end{tcolorbox}

\begin{tcolorbox}[promptbox,title={Prompt for LLM-Judge Scoring},fonttitle=\small]
You are evaluating an AI-generated planning output for a personal goal.

The input includes the raw system output and a structured extracted\_evidence record. Use the extracted\_evidence as an audit aid, not a replacement for reading the raw output. If the raw output and extracted evidence conflict, trust the raw output and mention the conflict in criterion evidence when relevant.

Score the output from 1 to 7 on each criterion, matching the human study Likert scale. \\
1 = very poor, harmful, or unusable. \\
2 = weak. \\
3 = partial but generic or incomplete. \\
4 = adequate. \\
5 = good and useful. \\
6 = strong, specific, and actionable. \\
7 = excellent, highly actionable, and well grounded. \\

Important: \\
- Longer is not automatically better. \\
- Reward relevant personalization, not verbosity. \\
- Penalize contradictions with user-provided context. \\
- Penalize unsupported assumptions. \\
- Penalize generic advice when the goal requires concrete artifacts. \\
- Evidence should be concise and cite concrete details from the output. \\
- Use extracted\_evidence to notice tangible outputs, context use, unresolved placeholders, unsupported details, workflow progress, context reuse, and user burden. Apply the same standard to every condition. \\

Criteria: \\
- plan\_quality: Does the output produce a coherent, useful plan for the user's goal? \\
- tangible\_result\_quality: Does it include directly usable artifacts such as drafts, schedules, checklists, templates, trackers, surveys, messages, or decision aids? \\
- confidence\_support: Would this output make the user more confident about taking real-world action? \\
- personalization\_context\_grounding: Does it use the user's stated context correctly and avoid ignoring important constraints? \\
- user\_burden\_reduction: Does it reduce the amount of steering, re-explaining, manual context management, and cleanup the user must do? \\

Match the real study protocol: \\
- Users had 25 minutes to create a plan and tangible results. \\
- Reward outputs that help the user get started, not outputs that merely describe what could be done. \\
- If the conversation is messy or incomplete, score the artifact actually produced, not an idealized answer. \\
- Do not reward verbosity by itself; an output can be long and still poor if it is generic, repetitive, or hard to act on. \\
- For workflow summaries, treat task trees, saved contexts, selected/reused contexts, and working drafts as part of the artifact's value. \\

Return only strict JSON matching the provided schema.

\end{tcolorbox}

\subsection{Validation}
\label{sec:appendix-judge-validation}

We validate the LLM judge using the within-subject JumpStarter--ChatGPT pairs from the human study. 
For each de-identified system output, the judge scores the same artifacts shown to human raters, and we compare its scores against human ratings of plan quality, tangible-result quality, and confidence to act on a 1--7 scale.

We assess judge validity using three pre-specified checks: (i) Spearman correlation with the human-rated \texttt{overall} construct, (ii) mean absolute error (MAE) on the 1--7 scale, and (iii) whether the \texttt{study\_quality} composite recovers the human-study winner within each JumpStarter--ChatGPT pair.

The judge passes all validation gates. 
It achieves \texttt{spearman\_overall}\,=\,0.779, exceeding our threshold of $\rho \geq 0.7$, and \texttt{mae\_overall}\,=\,1.064, approximately within one Likert point. The mean within-pair margin is also positive (\texttt{study\_quality\_mean\_margin}\,=\,1.05), indicating that the judge consistently favors the same condition as the human-study ratings.
These results suggest that the judge's scores are well aligned with human ratings and preserve the direction of human-study preferences.





\section{Detailed Baseline Methods Descriptions}
\label{app:details_baseline}

We compare JumpStarter against the following baseline conditions.

\paragraph{ChatGPT baselines.}
We include three ChatGPT baselines using \texttt{GPT-4o} in a multi-turn planning setting, mirroring the user study condition in which participants planned with the ChatGPT interface. In \textit{vanilla ChatGPT}, the simulated user directly interacts with the model to plan the goal. In \textit{ChatGPT + elicited context}, the simulated user first answers global context questions before beginning the planning conversation. In \textit{ChatGPT + structured summary}, the elicited context is organized into a structured summary and provided to the model before planning begins. We also include \textit{single-turn decomposition}, which receives the same elicited context and is explicitly prompted to decompose the goal into subtasks and draft each one in a single turn, without an external task structure, per-subtask context selection, or reuse.

\paragraph{Single-turn decomposition results.}
\label{app:single-turn}
Explicitly prompting \texttt{GPT-4o} to decompose the goal and draft every subtask in a single turn raises the score from 3.70 (ChatGPT + elicited context) to 4.05, yet \textit{JumpStarter-Shallow} still outperforms this single-turn decomposition baseline (+0.22; Table~\ref{tab:auto-results}). The advantage lies in context curation and reuse (+1.36 on the judge's 1--7 scale) and personalization (+0.66), while decomposition quality is similar (+0.12), suggesting that \texttt{GPT-4o} decomposes goals well on its own but cannot attach the right context to each subtask in a single turn.

\paragraph{Planning and memory-agent baselines.}
We include representative baselines that capture alternative agentic approaches to complex planning. \textit{ADaPT} \citep{prasad2024adapt} represents recursive task decomposition for LLM planning. \textit{Ask-Before-Plan} \citep{zhang2024ask} represents clarification-first planning, where the model asks questions before generating a plan. We also include an unstructured memory-RAG baseline inspired by memory-augmented agents such as MemGPT \citep{packer2023memgpt}, where relevant prior context is retrieved from a flat memory store and provided to the model during planning. 
Finally, the \textit{integrated agentic planner} combines these capabilities in one loop: it runs \textit{JumpStarter}'s global elicitation step, retrieves from a memory store built from the user's materials, queries a simulated browse tool that returns generic guidance for the goal's deliverables, decomposes the goal with \textit{JumpStarter}'s subtask-generation prompt, and selects up to five relevant context items for each subtask before drafting. Unlike \textit{JumpStarter}, it neither elicits context for individual subtasks nor reuses drafts across subtasks.
These baselines test whether \textit{JumpStarter}'s gains can be achieved by decomposition, clarification, retrieval, or their combination, rather than by organizing elicited and reused context around task structure.

\paragraph{Integrated agentic planner results.}
\label{app:integrated-agent}
The integrated agentic planner scores 3.94. \textit{JumpStarter-Shallow} outperforms it by +0.33 (95\% CI [0.28, 0.38]), and \textit{JumpStarter-Recursive}, rerun alongside the agent, by +0.26 [0.20, 0.31]; the rerun matches the main run (difference $-0.004$ [$-0.06$, 0.05]). The agent is not short of content: it writes as many words as \textit{JumpStarter-Recursive} (1,100 vs.\ 1,071) and ties it on tangible-result quality and completeness. Its deficit is concentrated in context curation and reuse (+0.58 for \textit{JumpStarter-Shallow}, +0.73 for \textit{JumpStarter-Recursive}) and in personalization and context grounding (+0.63, +0.57): it has the user's elicited context but does not attach it to the right subtask or carry it forward. In the traces, the agent never routes a saved draft into a later subtask (0 reuse events per session, vs.\ 1.48 for \textit{JumpStarter}) and elicits less context (3.0 items per session, vs.\ 6.96), since it asks only the global questions. \textit{JumpStarter-Recursive}'s margin over the agent is largest on multi-track goals (+0.32, vs.\ +0.23 on high- and +0.24 on medium-complexity goals), where subtasks depend on each other most.

\paragraph{Workflow ablations and context-control baselines.}
We include workflow ablations to isolate the contribution of each component in task-structured context curation. Each ablation removes or replaces one mechanism of the \textit{JumpStarter} workflow with recursive decomposition enabled. \textit{Random context selection} replaces task-local selection with a randomly sampled context subset, testing whether any context reduction is sufficient. \textit{No context selection} removes the selection step and passes all available context to every subtask, testing whether performance improves simply by exposing more information. \textit{No context reuse} prevents saved drafts and intermediate artifacts from being carried forward, and \textit{no elicitation} removes proactive context gathering. Finally, \textit{JumpStarter-Recursive} uses the same context-curation mechanisms as \textit{JumpStarter-Shallow} but allows subtasks to be decomposed further. Together, these conditions test whether gains come from selecting relevant context, reusing intermediate artifacts, eliciting missing information, and using lightweight task structure rather than indiscriminately adding context or increasing decomposition depth.

\section{LLM Prompts}
\label{sec:prompts}

\subsection{Context Elicitation}
\label{app:prompt_ce}

\subsubsection{Goal Initialization and Global Context}
\label{app:prompt_ce_root}

\begin{figure}[ht]
\centering
    \includegraphics[width=\linewidth]{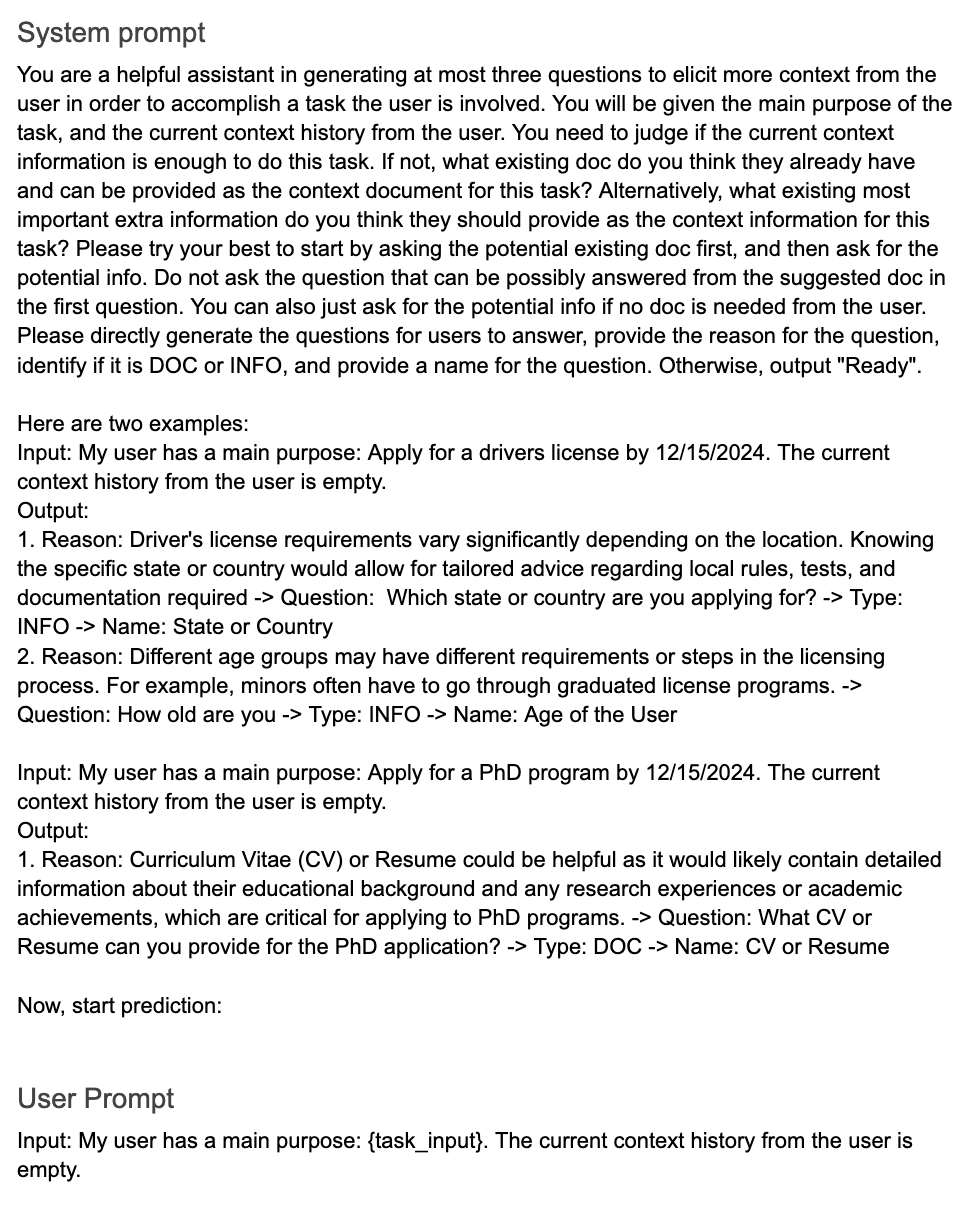}
    \caption{Prompts for Context Elicitation for Goal Input.}
    \label{app:p_ce_root}
\end{figure}

Figure \ref{app:p_ce_root} details the prompts used for eliciting global context for the goal initialization process introduced in Section \ref{sec:ce_goal}.

\subsubsection{Answer Draft Creation and Iteration}
\label{app:prompt_ce_draft}

\begin{figure}[h]
\centering
    \includegraphics[width=\linewidth]{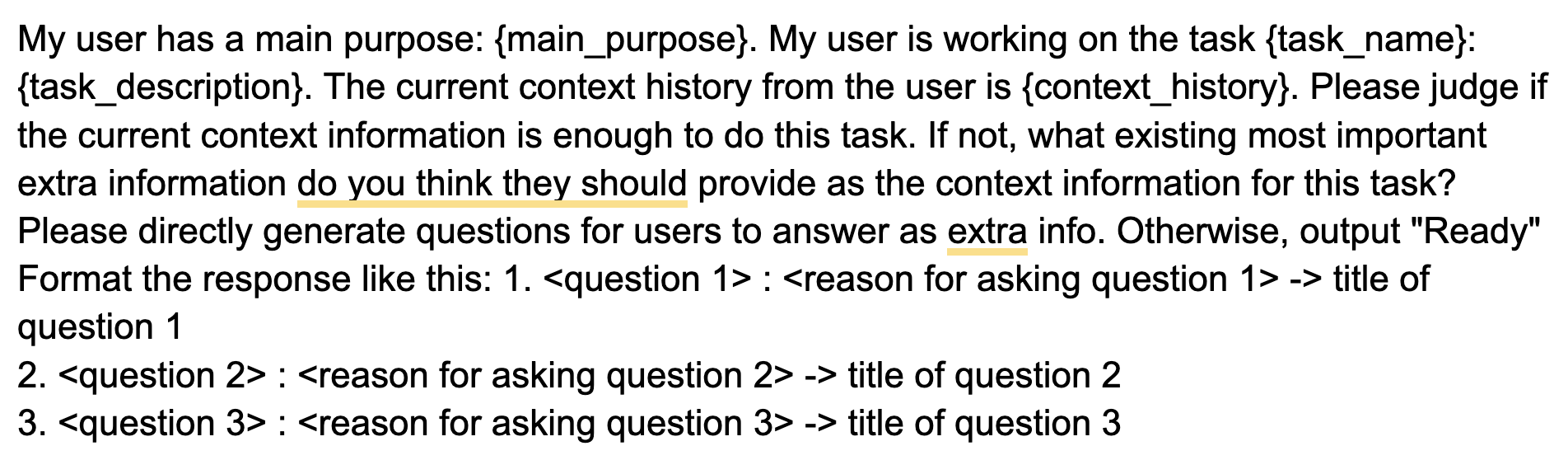}
    \caption{Prompts for Context Elicitation for Answer Draft Creation.}
    \label{app:p_ce_drafting}
\end{figure}

Figure \ref{app:p_ce_drafting} presents the prompts used to generate an answer draft for each subtask.

\subsection{Context Selection}
\label{app:prompt_cc}

\subsubsection{Answer Draft Generation}
\label{app:prompt_csd}

\begin{figure}[h]
\centering
    \includegraphics[width=\linewidth]{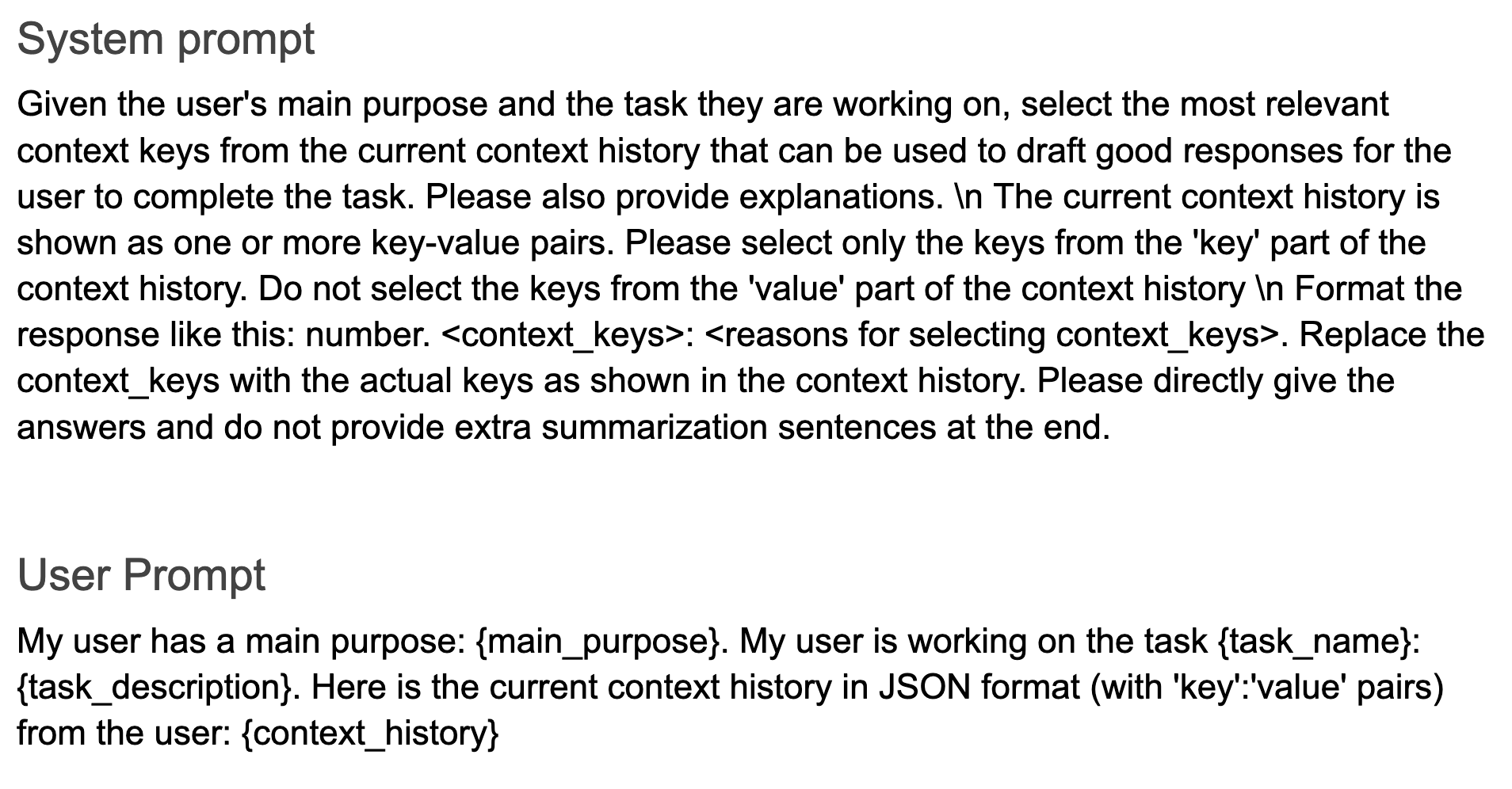}
    \caption{Prompts for Context Selection for Answer Draft Creation.}
    \label{app:p_cs_drafting}
\end{figure}

Figure \ref{app:p_cs_drafting} specifies the prompts used to select relevant context for generating an answer draft.

\subsubsection{Task Forking}
\label{app:p_cs_forksing}

The prompt to conduct the task forking is as follows: 
\begin{itemize}
    \item \textit{My user has a main purpose: \{main\_purpose\}. My user is working on the task \{task\_name\}: \{task\_description\}. My user needs to break down the task into sub-tasks. Here is the current context history from the user: \{context\_history\}. Please select the most relevant context key from the current context history that can be used to better decompose the current task into several sub-tasks for the user to get started. Do not help the user to break down the task. Please also provide explanations. Format the response like this: <context\_key>: <reasons>. Replace the context\_key with the actual key in the context history.}
\end{itemize}

\subsection{Task Decomposition}
\begin{figure}[ht]
\centering
    \includegraphics[width=\linewidth]{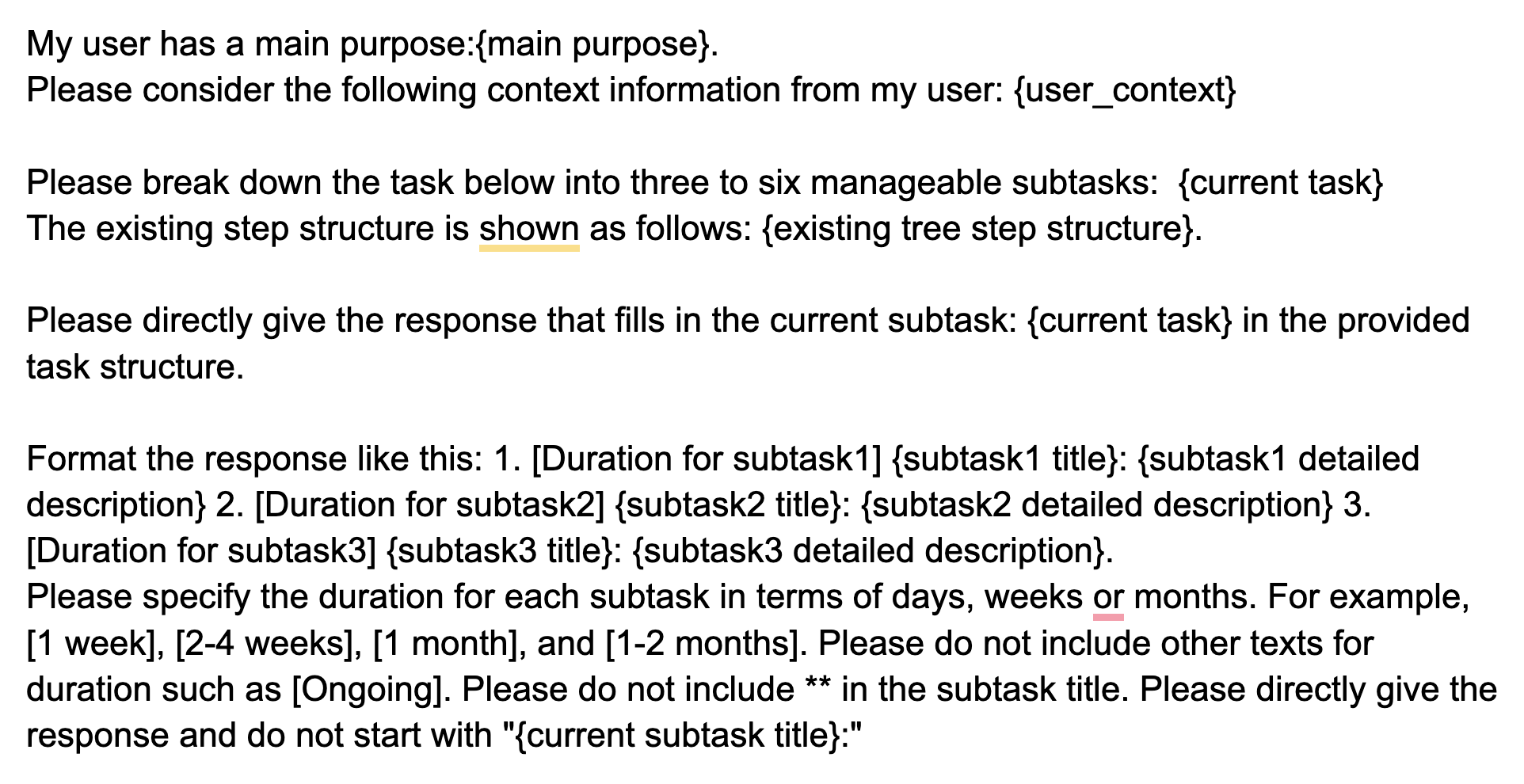}
    \caption{The prompt for Subtask Generation.}
    \label{app:p_sg}
\end{figure}

\begin{figure}[h]
\centering
    \includegraphics[width=\linewidth]{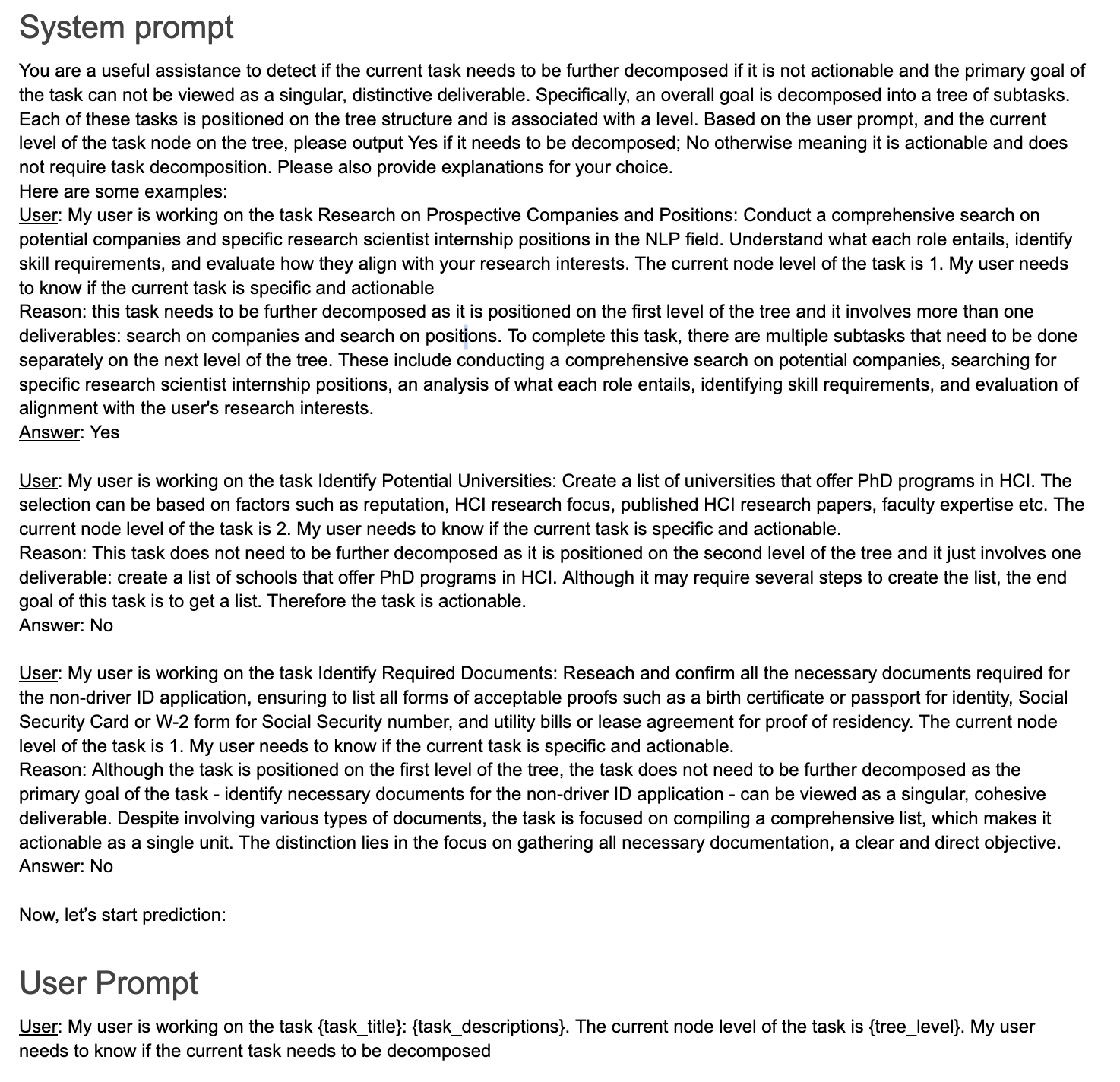}
    \caption{The prompt for Subtask Detection.}
    \label{app:p_sd}
\end{figure}

\begin{figure}[h]
\centering
    \includegraphics[width=\linewidth]{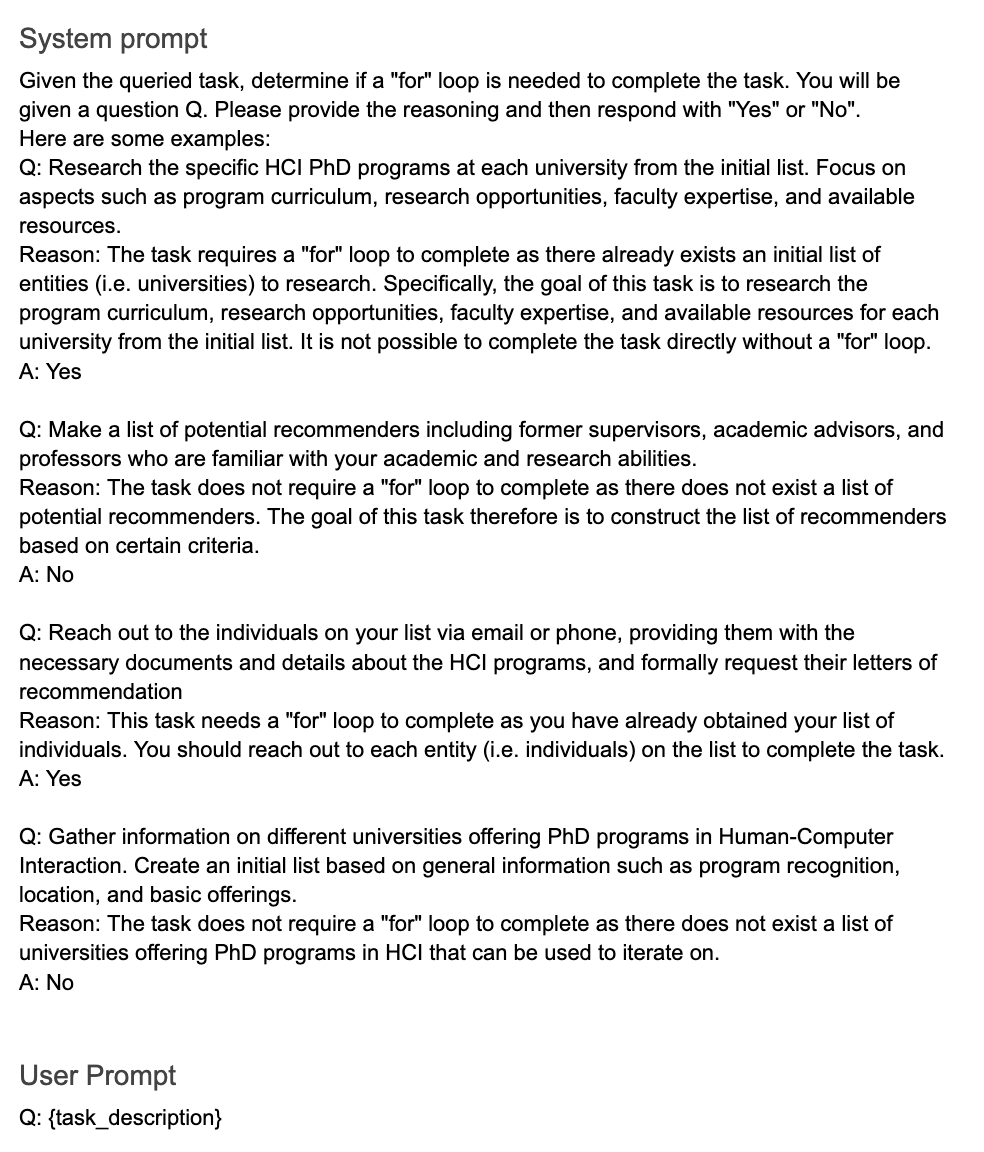}
    \caption{The prompt for Task Forking.}
    \label{app:p_forking}
\end{figure}

This section presents the prompts used for task decomposition, organized into three core components: Subtask Generation (Figure \ref{app:p_sg}), Subtask Detection (Figure \ref{app:p_sd}), and Task Forking (Figure \ref{app:p_forking}).



\subsection{Answer Draft Creation}
\label{app:prompt_iterate}

The prompt to generate the answer drafts is shown as follows:
\begin{itemize}
    \item \textit{My user has a main purpose: \{main purpose\}. Please consider the following context information from my user: \{user\_context\}. My user needs help with the current task \{current task\}: \{task description\}}
\end{itemize}

\subsection{Prompts for Technical Evaluation of Subtask Detection}
\label{app:prompt_techeval_sd}

\subsubsection{Zero-shot Prompting}
The prompt for zero-shot for the task of subtask detection is demonstrated below:
\begin{itemize}
    \item System prompt: \textit{You are a useful assistance to detect if the current task needs to be further decomposed if it is not actionable and the primary goal of the task can not be viewed as a singular, distinctive deliverable. Based on the user prompt, please output Yes if it needs to be decomposed; No otherwise meaning it is actionable and does not require task decomposition.}
    \item User Prompt: \textit{My user is working on the task \{task title\}: \{task description\}. My user needs to know if the current task needs to be decomposed.}
\end{itemize}

\subsubsection{Few-shot Prompting}

\begin{figure}[h]
\centering
    \includegraphics[width=\linewidth]{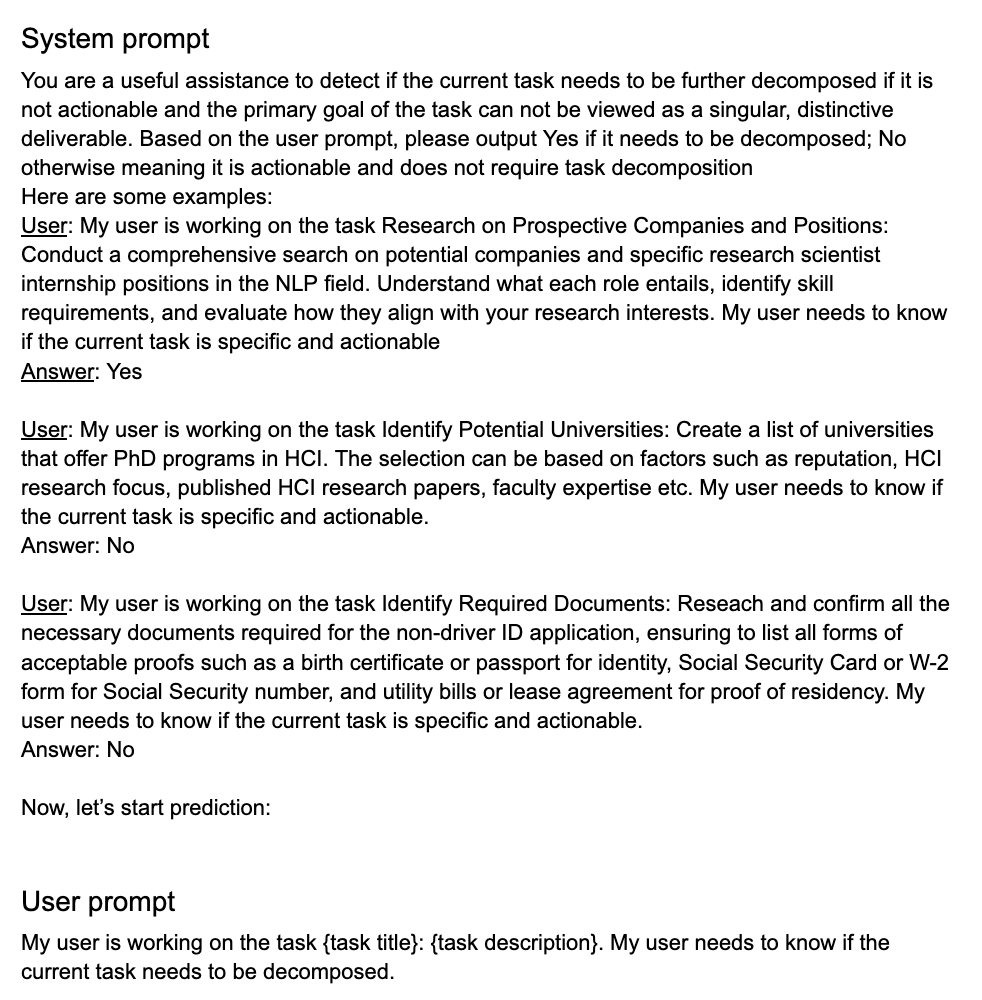}
    \caption{The few-shot-only prompt for Subtask Detection.}
    \label{app:p_few}
\end{figure}

The prompt for few-shot-only prompting is shown in Figure \ref{app:p_few}. Note that we used three in-context examples in the prompt.

\subsubsection{Few-shot + CoT}
\begin{figure}[t]
\centering
    \includegraphics[width=\linewidth]{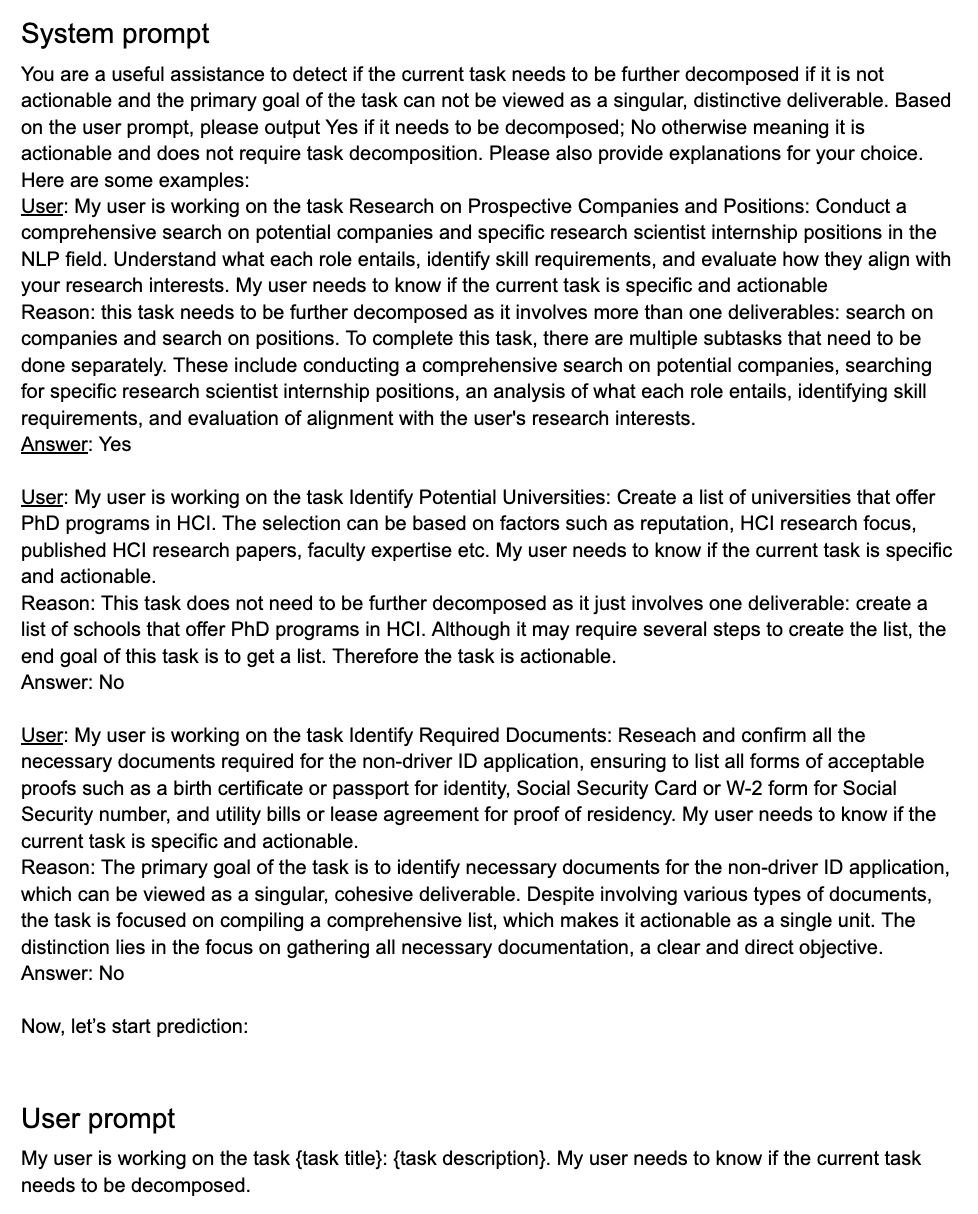}
    \caption{The prompt for few-shot + CoT for Subtask Detection.}
    \label{app:p_few_cot}
\end{figure}

We constructed the prompt in a Chain-of-Thought fashion, where GPT-4 is instructed to first generate the reasoning and then the answer. The prompt is shown in Figure \ref{app:p_few_cot}.

\subsubsection{Few-shot + CoT + Draft}
\begin{figure*}[h]
\centering
    \includegraphics[width=\linewidth]{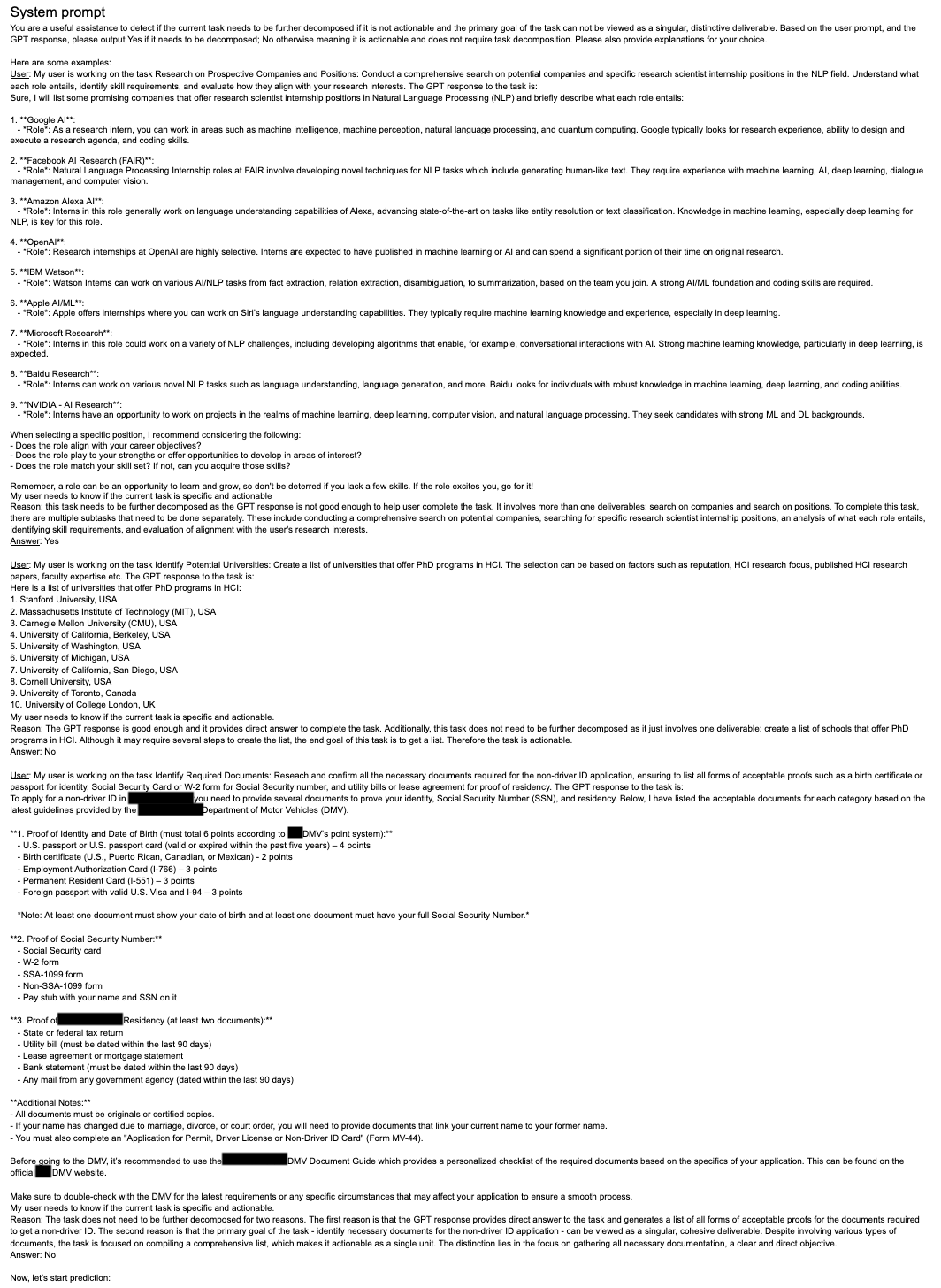}
    \caption{The prompt for few-shot + CoT + Draft for Subtask Detection.}
    \label{app:p_few_cot_draft}
\end{figure*}

We experimented with incorporating both CoT and the initial working solution draft into the prompt. The system prompt is shown in Figure \ref{app:p_few_cot_draft}. For the user prompt, before detecting subtasks. we first generated the initial working solution draft for the current task. The user prompt is shown below:
\begin{itemize}
    \item \textit{My user is working on the task \{task title\}: \{task description\}. The GPT response to the task is: \{Draft\}. My user needs to know if the current task is specific and actionable.}
\end{itemize}

\subsubsection{Few-shot + CoT + Tree + Draft}

\begin{figure*}[h]
\centering
    \includegraphics[width=\linewidth]{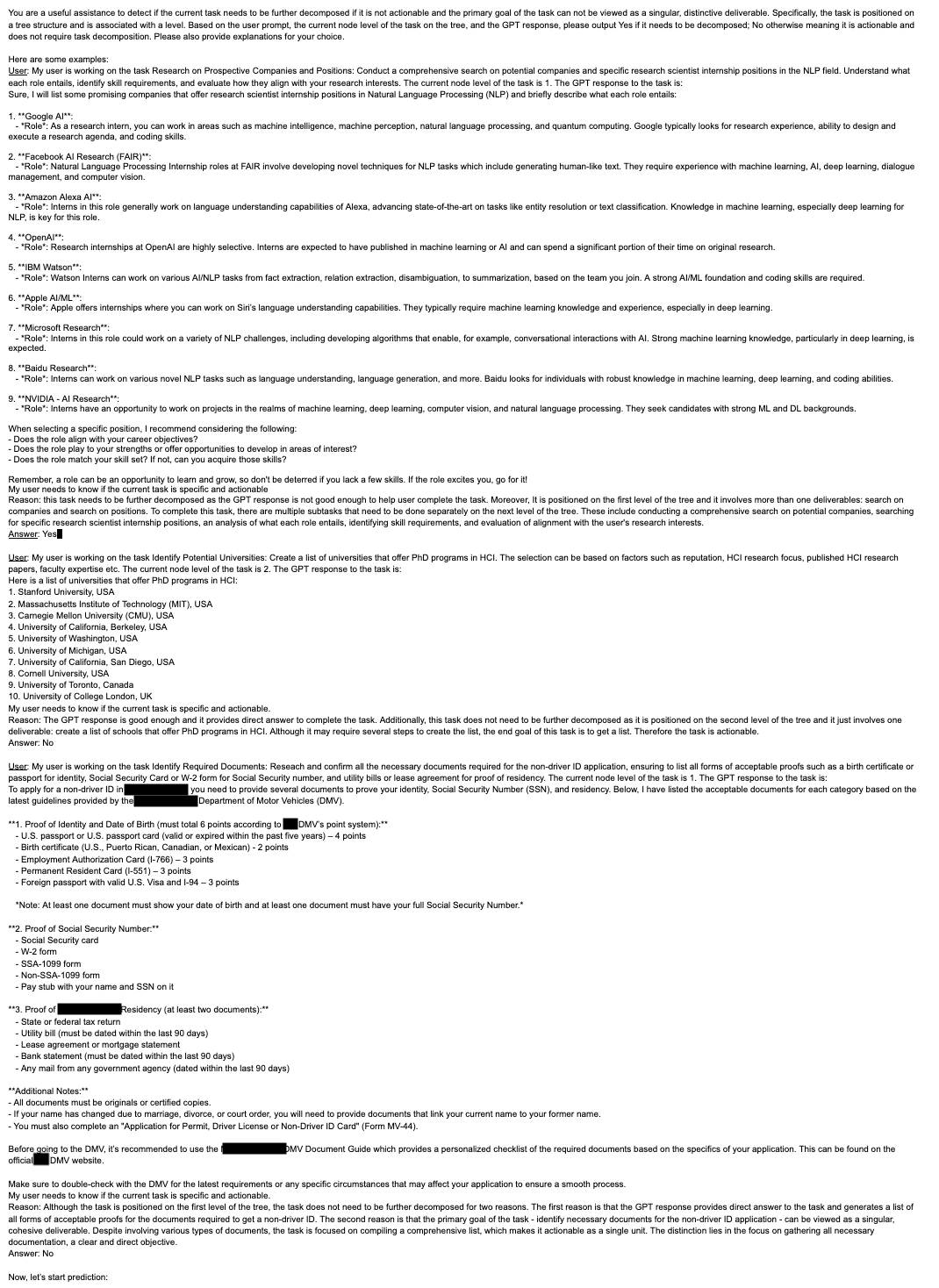}
    \caption{The prompt for few-shot + CoT + Tree + Draft for Subtask Detection.}
    \label{app:p_few_cot_draft_tree}
\end{figure*}

To construct the system prompt for this setting, we incorporate the tree level of each task into the prompt. The prompt is shown in Figure \ref{app:p_few_cot_draft_tree}. Additionally, for the current task at hand, its tree-level information is also presented in the user prompt, as shown below:
\begin{itemize}
    \item \textit{My user is working on the task \{task title\}: \{task description\}. The current node level of the task is \{level\}. The GPT response to the task is: \{Draft\}. My user needs to know if the current task is specific and actionable.}
\end{itemize}

\end{document}